\documentclass[twocolumn]{aastex701}
\usepackage{fontawesome}
\usepackage{CJK}
\usepackage{natbib}
\usepackage{amsmath}

\usepackage{xspace}
\usepackage[utf8]{inputenc}

\graphicspath{{figures}}

\newcommand{\teff}{$T_{\rm eff}$\xspace}
\newcommand{\logg}{$\log{(g)}$\xspace}
\newcommand{\vsini}{$v\sin{i}$\xspace}

\newcommand{\feh}{\rm{[Fe/H]}\xspace}
\newcommand{\monh}{\rm{[M/H]}\xspace}

\newcommand{\aonm}{$\rm{[\alpha/M]}$\xspace}

\newcommand{\mpp}{Metal Pipe\xspace}

\received{---}
\revised{---}
\accepted{---}
\submitjournal{AJ}

\shorttitle{Metal Pipe}
\shortauthors{Kolecki \& Weiss}

\graphicspath{{./}{}}

\begin{document}
\begin{CJK*}{UTF8}{gbsn}

\title{Metal Pipe: A Broadly-Applicable Stellar Abundance Pipeline Using Isochronal Parameters}

\author{Jared R. Kolecki}\affiliation{Department of Physics \& Astronomy, University of Notre Dame, Notre Dame, Indiana 46556, USA}\email{jkolecki@nd.edu}

\author{Lauren M. Weiss}\affiliation{Department of Physics \& Astronomy, University of Notre Dame, Notre Dame, Indiana 46556, USA}\email{lweiss4@nd.edu}

\correspondingauthor{Jared Kolecki}
\email{jkolecki@nd.edu}

\begin{abstract}

Characterizing exoplanet host stars at a population level requires a method of homogeneously characterizing stellar properties across all spectral types.  To this end, we have developed \mpp, a new code for determining stellar parameters and abundances, which is designed for use across a wider range of spectral types than many commonly used codes. It combines the widely-used package \texttt{MOOG} with photometric stellar parameters, a user-supplied high-resolution spectrum, and a newly curated list of spectral lines. \mpp outputs values for \teff, \logg, $M_*$, $R_*$, and $L_*$ from isochrones, and abundances of C, O, Na, Mg, Al, Si, S, Ca, Ti, and Fe from \texttt{MOOG}. In this paper, we describe the \mpp algorithm and tests of \mpp against previous abundance measurements on archival HIRES spectra of 503 F, G, and K type stars. We find RMS scatters of $\sim100\textrm{ K}$ in \teff, $\sim0.10 \textrm{ dex}$ in \logg, and $\sim0.10 \textrm{ dex}$ for all measured abundances. These values are comparable to estimated measurement uncertainties, verifying \mpp for continued use in building a detailed abundance catalog.  Future papers in this series will test \mpp's applicability to late K and M dwarf stars, and provide other improvements.
\end{abstract}


\section{Introduction}\label{Introduction}
In the search for patterns relating host star chemical abundances and planetary properties, a homogeneous abundance analysis of a large catalog of planet-hosting stars would be beneficial. However, the diversity of planet hosting star spectral types, wavelength ranges covered by spectrographs, and stellar atmospheric models makes it difficult to compile a homogeneous catalog from the literature \citep{Hinkel+2016}.  

As a potential solution, we have developed \mpp, a framework for determining the detailed chemical composition of stars. We built \mpp with the goal of characterizing FGK and M dwarfs with high resolution spectra, applying a consistent methodology regardless of spectral type.  This paper is the first in a series focused on the development, testing, and application of \mpp. We describe the algorithms for fitting stellar parameters and abundances, and benchmark \mpp on a set of nearby stars with high-quality literature data.

This introduction discusses the methods and challenges of obtaining stellar abundance measurements from high-resolution spectroscopy motivating the creation of \mpp. In Section \ref{Sec:StarPlanetChem}, we discuss how stellar composition can impact planet formation, motivating our choice of initial elemental abundances in \mpp.  We describe the dependencies of \mpp in Section \ref{Sec:Dependencies} and required inputs in Section \ref{Sec:Inputs}.  The \mpp algorithm (as it is currently implemented) is described in Section \ref{section:MPP}.  We compare the results of \mpp to previous work in Section \ref{Sec:Verification}.  We summarize and discuss possible future directions in Section \ref{Sec:Future}.

\subsection{Challenges of Abundance Catalogs of Planet Hosting Stars}\label{Sec:Challenges}
Methodological differences can have a significant impact on the final reported values of stellar parameters and abundances. This makes it difficult to directly compare stellar abundances reported by different catalogs. Measurements of the same stellar sample using different methods may show systematic offsets of 0.1-0.2 dex \citep[see e.g.][]{Hinkel+2016,Jofre+2018,BC2019}. 

Additionally, catalogs have historically been tuned to work for either FGK type stars \citep[e.g.][]{Adibekyan+2012,Brewer+2015} or M dwarfs \citep[][]{Mann+2015,Behmard+2025}. This distinction largely arises because visible-wavelength spectra of M dwarfs are significantly polluted with molecular lines, which blend with the atomic lines that are typically measured to determine the abundances of FGK stars.  Another challenge is that the atmosphere models that are typically used are missing physics (e.g, accurate line data, 3D hydrodynamic effects, in some cases non-LTE corrections), causing models to diverge from reality for stars that are increasingly different from the sun.  Therefore, it is rare for a single abundance catalog to include both classes of stars. This creates a discontinuity near the median temperature of planet hosts. The temperature ranges of the cited catalogs, and their other basic properties (wavelength range, spectral resolution, sample size) are described in Table \ref{table:cats}. 

\begin{deluxetable*}{lrrrrrrrr}[t]
\tablecaption{
    Basic statistics of various abundance catalogs \label{table:cats}
}
\tablehead{
    \colhead{Catalog} &
    \colhead{\teff Range} &
    \colhead{n Stars} &
    \colhead{n Elements} &
    \colhead{CHNOPS} &
    \colhead{Lithophiles} &
    \colhead{$\lambda$ Range (\AA)} & 
    \colhead{$\lambda/\delta\lambda$} &
    \colhead{S/N} 
    }

\startdata        
        \mpp (This Work)     & 4565 - 6449 & 503 & 10 & 3 & 6 & 3650 - 7950\phn & 70000 & $\sim$200 \\ 
        \mpp (Anticipated)   & 3200 - 6500 & $>$1000 & $\sim$15 & 4 - 5 & 6 & \xspace3650 - 25000 & varied & $\sim$100 \\ 
        \citet{Brewer+2016}    & 4702 - 6773 & 1615 & 15 & 3 & 6 & 3650 - 7950\phn & 70000 & $\sim$200 \\ 
        \citet{Adibekyan+2012} & 4487 - 7212 & 1111 & 12 & 0 & 6 & 3800 - 6900\phn & 110000 & $\sim$200 \\ 
        \citet{Mann+2015}      & 2700 - 4131 & 183 & 1 & 0 & 0  & \xspace3200 - 24000 & $\sim1500$ & $\sim$150 \\
\enddata
    
\end{deluxetable*}



\subsubsection{Importance of Low-Mass Stars}
Late-K and M-type main sequence stars offer profound opportunities to characterize Earth-sized exoplanets. They represent a disproportionate fraction of small planet discoveries, thanks to their relatively deep transits and large RV semi-amplitudes (as compared to similar planets around F and G type stars, \citealt{Winn2010, Lovis2010}). The NASA Exoplanet Archive\footnote{accessed 2025-07-24} lists 1463 planets with $R_{\rm{p}} < 1.8R_{\oplus}$. Of these, 390 ($26\%$) orbit stars with $T_{\rm{eff}} < 4700\rm{K}$. Of planets with $R_{\rm{p}} > 1.8R_{\oplus}$, 686 out of 4437 ($15\%$) meet this same criterion. Abundance measurements of low-mass stars are therefore essential to understanding the formation chemistry of terrestrial exoplanets.

\subsubsection{Bridging The Gap To M Dwarfs}
Previous efforts to generate catalogs of homogeneously determined stellar parameters for planet hosting stars have had moderate success, but with the caveat that bridging the gap between FGK and M type stars remains a challenge.  The California-Kepler Survey (CKS) was a homogeneous catalog of 1307 host stars of 2025 Kepler-detected exoplanets around F, G, and early K type stars \citep{Petigura+2017}.  This catalog used the code \texttt{specmatch} to obtain \teff, \logg, \vsini, and \feh based on low-SNR ($\sim10$), high-resolution ($\sim60,000$) spectra of the stars \citep{Petigura2015}.  However, the accuracy of \texttt{specmatch} degraded significantly for stellar temperatures below $\sim4600$ K.  The workaround was the development of a companion code, \texttt{specmatch-emp}, which assigned stellar properties based on empricial comparison to a library of cool stars with known properties \citep{Yee2017}.  

The development of \texttt{specmatch-emp} allowed a new campaign, CKS-cool, which extended the CKS catalog to late K and M dwarf planet hosts \citep{Petigura2022}.  While these catalogs represent a notable achievement in each giving homogeneous stellar properties, the gap between them remains a possible source of systematics when comparing stars above vs. below 4600 K. In particular, comparing abundances of elements other than iron across this gap has not yet been shown to be reliable. This is partly due to the limitations of stellar atmosphere models, which become less accurate with decreasing effective temperatures, as well as to the challenges of identifying atomic lines in the optical spectra of late-K and M stars.

By taking into consideration the particular needs of cool stars with carefully chosen models and methods, and implementing a methodology which can be applied uniformly on either side of the 4600K temperature divide, we have developed \mpp with the intent to bridge this gap.

\subsubsection{Optical vs. NIR Spectra}
One strategy the exoplanet community plans to employ for bridging the 4600K gap is the use of near-infrared (NIR) spectra. Stars cooler than 4600K are notorious for having significant molecular absorption bands in their spectra. These bands cover almost the entire optical wavelength range in a dense ``forest'' of molecular lines. These thousands of lines are extremely challenging to model with sufficient precision to perform abundance analyses on late-K and M dwarfs using their optical spectra.

Recent advancements in detector technology enabled a new generation of high-resolution (NIR) spectrographs (e.g. SPIRou, \citealt{Artigau+2014}; IGRINS, \citealt{IGRINS}). With these instruments, it is possible to observe cool stars at wavelengths where molecular bands are still present, but significantly weaker than in the optical regime. In a sense, the molecular line forest becomes ``thinner'' at NIR wavelengths. This allows for clearer identification and more accurate modeling of line features of interest for abundance analysis.

\subsubsection{Stellar Parameters}
Historically, it has been difficult to derive accurate bulk physical properties (\teff and \logg) values for cool stars. In FGK stars, these are often determined spectroscopically, by fitting a synthetic spectrum to strong spectral lines such as the Na-D doublet, Mg-b triplet, or the H-$\alpha$ and H-$\beta$ lines \citep[e.g.][]{West+2009,Brewer+2015}. Other methods include enforcing ionization balance and excitation equilibrium of iron \citep[e.g.][]{Adibekyan+2012}. However, in M dwarfs, these strong spectral lines in the optical wavelength regime are too obscured by molecular absorption to be precisely measured. Additionally, low-mass stars are not hot enough to ionize iron. Thus, the traditional spectroscopic methods of deriving \teff and \logg for FGK stars simply do not work for late-K and M dwarfs.

Fortunately, new parallaxes from the Gaia mission \citep{Gaia+2016} allow for the use of another technique, which applies equally well for all stars\footnote{excluding young stars with significant IR excess}: photometric SED fitting. As of Data Release 3 \citep{GaiaDR3}, Gaia has measured the parallaxes of over 1.5 billion celestial objects. With these measurements, it is possible to convert a star's apparent magnitude into an absolute brightness with greater precision than was possible in the pre-Gaia era.

Having made the conversion to absolute magnitudes, we can compare the brightness of a star in various photometric passbands to synthetic models computed from stellar isochrones \citep[e.g. MIST,][]{Choi+2016}. Comparing the observed photometry to these models allows for a precise determination of a star's \teff, mass, radius, and luminosity using photometry alone. While this approach does not fully eliminate the potential for systematic effects to influence our results, as such effects may arise in the computation of the MIST synthetic photometry, it does allow us to determine stellar parameters for M stars using the same methods we use for FGK stars, making our pipeline more broadly applicable, and our results more self-consistent between spectral types, than has been traditionally available from spectroscopic methods of determining stellar parameters.


\section{Stellar Abundances and Planet Formation}\label{Sec:StarPlanetChem}
Our choice of which elements to include in the initial version of \mpp stems from the basic tenets of star and planet formation theories. A star begins its life as a small, dense core of material \citep[$r \lesssim 0.05$\xspace pc, see][]{Pineda+2023} in a molecular cloud. These cores are much smaller than typical structures in molecular clouds \citep[$l \sim1$\xspace pc, see e.g.][and references therein]{Dib+2020}. From this, we can assume any given pre-stellar core has an approximately uniform composition.

Cores which reach a critical density will collapse, forming a protostar and protoplanetary disk \citep[see][for a review]{Pineda+2023}. Because the protostar and its disk form from the same core of material, we may assume that their compositions are closely linked. This compositional link is maintained throughout the life of the system, meaning fully formed planetary systems should also mimic the bulk composition of their host star.

\subsection{Compositional Variations and Geochemical Theory}
Stars exhibit compositional variations dependent on when and where they formed in the galaxy \citep[e.g. alpha enhancement, see][]{Edvardsson+1993,Hawkins+2015}. To first order, we can assume that their protoplanetary disks also exhibited similar variations.  However, various processes can alter chemistry in the protoplanetary disk as a function of both position in the disk and time (see \citealt{Bergin2007} for a review).  The extent to which alterations in disk chemistry correlates with physical and chemical mechanisms inherent in planet formation is not yet well understood.

Some abundances of interest are C, O, Na, Mg, Al, Si, S, Ca, Ti, and Fe. Taken together, these elements constitute over 97\% of Earth's solid mass \citep{Wang+2018}. As a consequence, they are critical tracers of planet formation processes. Aside from iron, these elements can be classified into two main groups according to their chemical role in planet formation and evolution.

\subsubsection{Lithophiles}
Lithophile (``rock-loving'') elements are the primary constituents of rocky planet mantles \citep{White2018}. The most chemically abundant of these species are Mg and Si. Estimates of planetary Fe/Mg and Mg/Si ratios suggest a strong connection between rocky planet and host star compositions \citep{Schulze+2021,Brinkman+2024}. Thus, determining the values of these ratios in a star can inform compositional analyses of its planets.

Other important lithophiles include Na, Al, Ca, and Ti. These elements mix together with O, Mg, and Si to create minerals of varying properties and complexity. Among the most significant of these properties is ``oxygen fugacity\footnote{The word ``fugacity'' derives from the Latin word \textit{fugere} (``to escape''). Thus, fugacity can best be understood as a ``tendency to escape.''}'' ($f\textrm{O}_2$).  The fugacity of oxygen (i.e. its tendency to escape a planet's mantle and seep into its atmosphere) has major implications for the chemistry of these layers in terrestrial and sub-Neptune planets \citep[see e.g.][]{Lichtenberg+2023}.

\subsubsection{CHNOPS}

Carbon, hydrogen, nitrogen, oxygen, phosphorus, and sulfur (collectively, ``CHNOPS'') are essential building blocks of life on Earth \citep[e.g.][]{Krijt+2023}. In addition, they also form the majority of atmospheric chemical species in planets of any size. As such, various literature studies \citep[e.g.][and references therein]{Turrini+2021,Pacetti+2022} explore the role of CHNOPS elements in uncovering the formation pathways of giant planets.

\subsection{Previous Spectroscopic Surveys of Planet Hosts}
\citet{Fischer+2005} found that metal-rich stars are more likely to host giant planets compared to metal-poor stars. The terms ``metal rich'' and ``metal poor'' refer to a star's bulk metallicity ([M/H]). [Fe/H] is commonly used as a proxy for [M/H], but this is only valid for stars with scaled-solar abundance patterns. Since the publication of this discovery, numerous other works have uncovered similar correlations of planet occurrence with bulk metallicity \citep[see e.g.][]{Xie+2016,Brewer+2018,Dong+2018,Petigura+2018}.

While bulk metallicity is straightforward to measure and has yielded interesting patterns relating to planetary properties, it is also the lowest-order approximation of a star's composition. Thus, it is unlikely to paint a complete portrait of the chemistry of planet formation. Previous work has cataloged the abundances of non-iron elements for planet-hosting stars \citep[e.g.][]{Brewer+2016, Brewer2018_kois}, but has not yet revealed strong correlations with planet architectures. 


\section{Dependencies and Input Models}\label{Sec:Dependencies}
Having established our motivation in writing \mpp and our choice of elements, we now describe the dependencies required for running \mpp.  \mpp is a C++17 framework built around the \texttt{nov2019} version of \texttt{MOOG}\footnote{retrieved from \url{https://www.as.utexas.edu/~chris/moog.html}} \citep{Sneden1973}. Note that \texttt{MOOG} solves the equations of stellar structure in one dimension under the assumption of local thermodynamic equilibrium (1DLTE). Although some precomputed non-LTE corrections are applied at the end (see Section \ref{section:nlte}), this means that \mpp is subject to the same limitations as other 1DLTE abundance analysis codes.

\mpp uses the minimization and interpolation modules of the GNU Scientific Library \citep[GSL, ][]{Galassi_2009}. \mpp uses additional Python 3 code to calculate stellar parameters using photometric data accessed from SIMBAD \citep{Wenger+2000} using \texttt{astroquery} \citep{Ginsburg+2019}, \texttt{astropy} \citep{Astropy13,Astropy18,Astropy}, and \texttt{scipy} \citep{Virtanen+2020}.

\mpp requires a number of input files. Some of these are editable or replaceable by the end user. Here, we discuss the required inputs and the level of customization each one offers. Full documentation is available at the \mpp Github repository\footnote{\url{https://github.com/kolecki4/Metal-Pipe}}.

\subsection{Stellar Spectrum}
The stellar spectrum should be in plain text format. Each line of text should include wavelength, flux, and uncertainty data, displayed in that order, separated by a single space. The stellar spectrum should also be normalized such that the continuum flux level is set to 1 at all wavelengths. Our particular algorithm for spectrum normalization is discussed in Section \ref{Spectra}.

This format ensures \mpp is agnostic to the choice of spectrograph used, so long as the data has sufficient resolution ($R \equiv \frac{\lambda}{\delta\lambda}$) and signal-to-noise ratio (S/N). As of this work, we recommend that any input stellar spectrum have $\textrm{S/N} \gtrsim 70$ and $R \gtrsim 45,000$. While \mpp can theoretically be run with any spectrum, its performance on spectra which do not meet these criteria has not been tested extensively and may be subpar.

\subsection{Line Lists}
\subsubsection{General Line List Data}
The line lists stored in the \texttt{mooglists} directory are supplied by \texttt{linemake}\footnote{\url{https://github.com/vmplacco/linemake}} \citep{Placco+2021} and stored in plain text format. If the user wishes to update the line data for a given transition, this data should be added to the file \texttt{mooglists/goodgf} for atomic transitions, or to the appropriate molecular data file for molecular transitions.

\subsubsection{Line Lists For Abundances}
\mpp allows the user to specify the list of spectral lines from which it will derive abundances. These lines are read from user-editable files located in the \texttt{linelists} directory. It is important to note that the wavelength data in these files must exactly match the wavelength data for the line contained in the \texttt{mooglists} directory.


\subsection{Isochrones}
Isochrones were sourced from the MIST database\footnote{\url{https://waps.cfa.harvard.edu/MIST/model_grids.html}} \citep{Choi+2016}. It is possible to substitute the included isochrone models with ones supplied by the user. However, if the user wishes to avoid modifying the source code of \mpp, the files supplied by the user must be named and formatted identically to those included.

\subsection{Model Atmosphere Grid}
Similarly to the isochrone grid, the model atmosphere grid can be substituted with one supplied by the user, by either modifying the source code, or by keeping file names and formats identical while replacing the data contained within them.

The included model atmosphere grid was calculated by \citet{Husser+2013} using the Phoenix stellar modeling code. We reformatted the files provided in the Goettingen Spectral Library\footnote{\url{https://phoenix.astro.physik.uni-goettingen.de/}} to conform to the ``GENERIC'' format as described by the \texttt{MOOG} documentation\footnote{\url{https://www.as.utexas.edu/~chris/codes/WRITEnov2019.pdf}}.

\section{Choosing Inputs}\label{Sec:Inputs}

\subsection{Line List Curation}\label{linelistcuration}
We use \texttt{linemake} \citep{Placco+2021} as the primary source of line data for \mpp, with additional data sourced from the Vienna Atomic Line Database (VALD, \citealt{Piskunov1995}). We use every line for every atomic species in the \texttt{linemake} database to generate our synthetic spectra. However, only a limited subset of these lines meet our criteria for use in abundance calculations: having minimal blended line features and having accurate transition data. We chose to curate lists of high-quality lines for \mpp to use for fitting abundances. 

For each species with a curated line list, we ran \mpp using every line with data in \texttt{linemake} on a small set ($n\approx10$) of stellar spectra obtained with the W. M. Keck Observatory's Keck Planet Finder \citep{Gibson+2016}. We visually inspected the line fits for each star, ensuring that each line was clearly distinguishable in a majority of our test spectra. 

\begin{figure*}
    \centering
    \includegraphics[width=\linewidth]{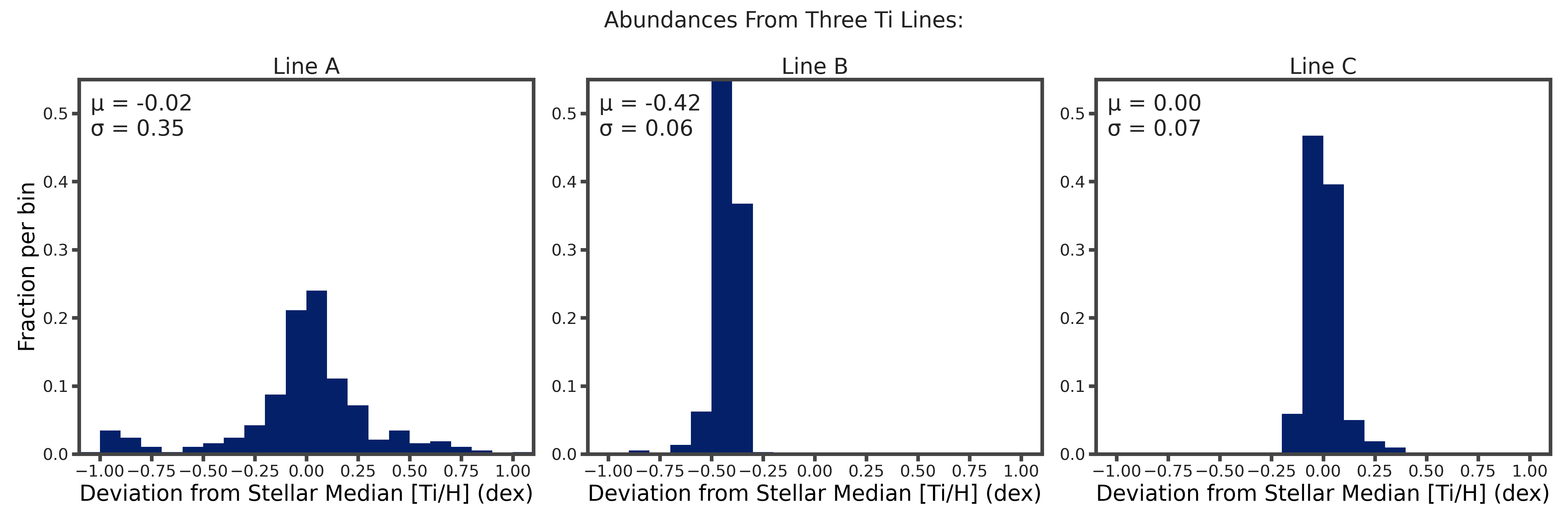}
    \caption{Histogram of abundances from three Ti lines as outlined in Section \ref{linelistcuration}. Line `A' returns highly variable abundance readings (note the broad distribution of abundance measurements). Line `B' achieves consistent abundance readings, but with a severe systematic offset from the median (i.e. this line consistently underestimates the Ti abundance), making it unsuitable for analysis. Line `C' is ideal for our work, achieving abundance values within 0.1 dex of the median in roughly 90\% of stars.}
    \label{fig:HistGallery}
\end{figure*}

Using only those lines which passed initial visual inspection, we ran \mpp on a larger subset ($n\approx250$) of stellar spectra. Once this was done, we plotted histograms for each line of its abundance results compared to the median derived from all lines. Through visual inspection of these histograms, we removed lines which consistently over- or under-estimated abundances. Lines that returned highly variable abundance readings were also removed. Representative histograms illustrating these cases, along with an example of a ``good'' line, are shown in Figure \ref{fig:HistGallery}.

After these quality cuts, we are left with line lists for each element which vary in length from $>400$ lines in the case of iron, to as few as 3 lines for oxygen.  The complete curated line lists are available in the \texttt{linelists} directory of \mpp\footnote{\url{https://github.com/kolecki4/Metal-Pipe}}.

\subsection{Spectra}\label{Spectra}

Reduced ``1-dimensional'' data products from a high resolution echelle spectrograph often contain large-scale fluctuations in the continuum. These fluctuations may correspond to spectrograph response, or to the star's pseudo-blackbody emission curve. To fit model line profiles to an observed spectrum, we must first normalize the observed spectrum. We define a ``normalized'' spectrum as one in which the continuum flux level is set to 1 at all wavelengths. 

To normalize our spectra, we apply data filtering techniques using SciPy \citep{Virtanen+2020} to derive a continuum directly from the observed spectral data. First, a median filter and maximum filter are applied to the 1D spectrum. This process identifies good ``continuum points,'' the anchor points in between spectral lines on which the continuum will be constructed. We apply a morphological closing filter to fill the majority of gaps in between continuum points. To completely remove wider spectral features (e.g. H-$\alpha$, other Fraunhofer lines, molecular bands), a convex hull \citep[see][]{Barber+1996} is applied over gaps $\geq 5\textrm{ \AA}$ in between continuum points. Finally, a smoothing filter is applied to remove any sudden ``jumps'' in the continuum.

\begin{figure}
    \centering
    \includegraphics[width=\linewidth]{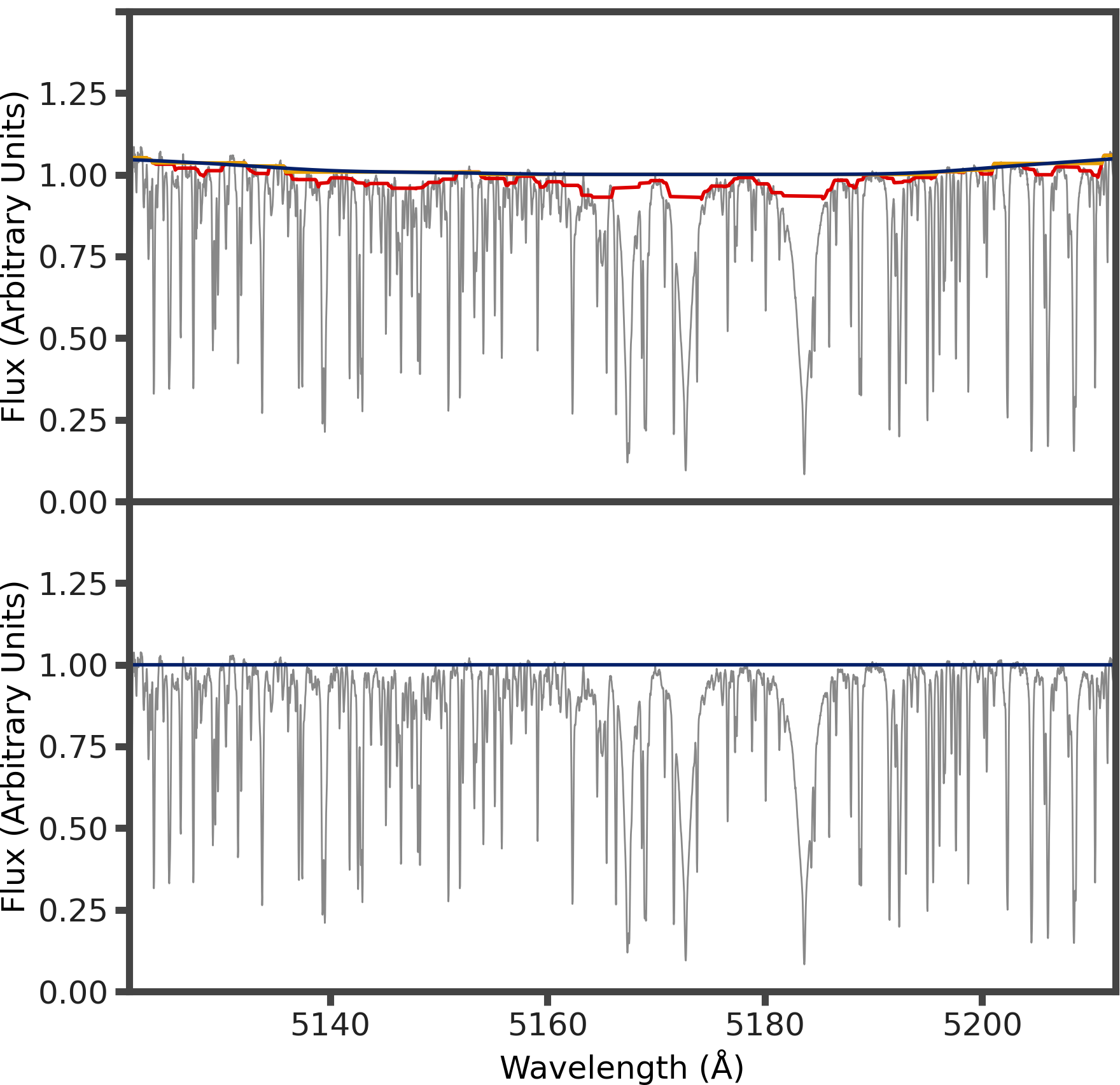}
    \caption{An illustration of our spectrum normalization process. \textbf{Top:} A 1-D flat-fielded HIRES spectrum of 51 Peg where flux has been scaled by an arbitrary constant. The red line shows the computed continuum after the median and maximum filters have been applied. The yellow line shows the result after the alpha ball and convex hull have been applied, and the blue line shows the final smoothed continuum. \textbf{Bottom:} The final continuum has been divided out. The blue line is set to 1 at all wavelengths.}
    \label{fig:Normalization}
\end{figure}

Once the continuum is constructed, we divide the original spectrum by it to create the normalized spectrum file used by \mpp. An illustration of the normalization process is shown in Figure \ref{fig:Normalization}.

\begin{figure*}[h!]
    \centering
    \includegraphics[width=\linewidth]{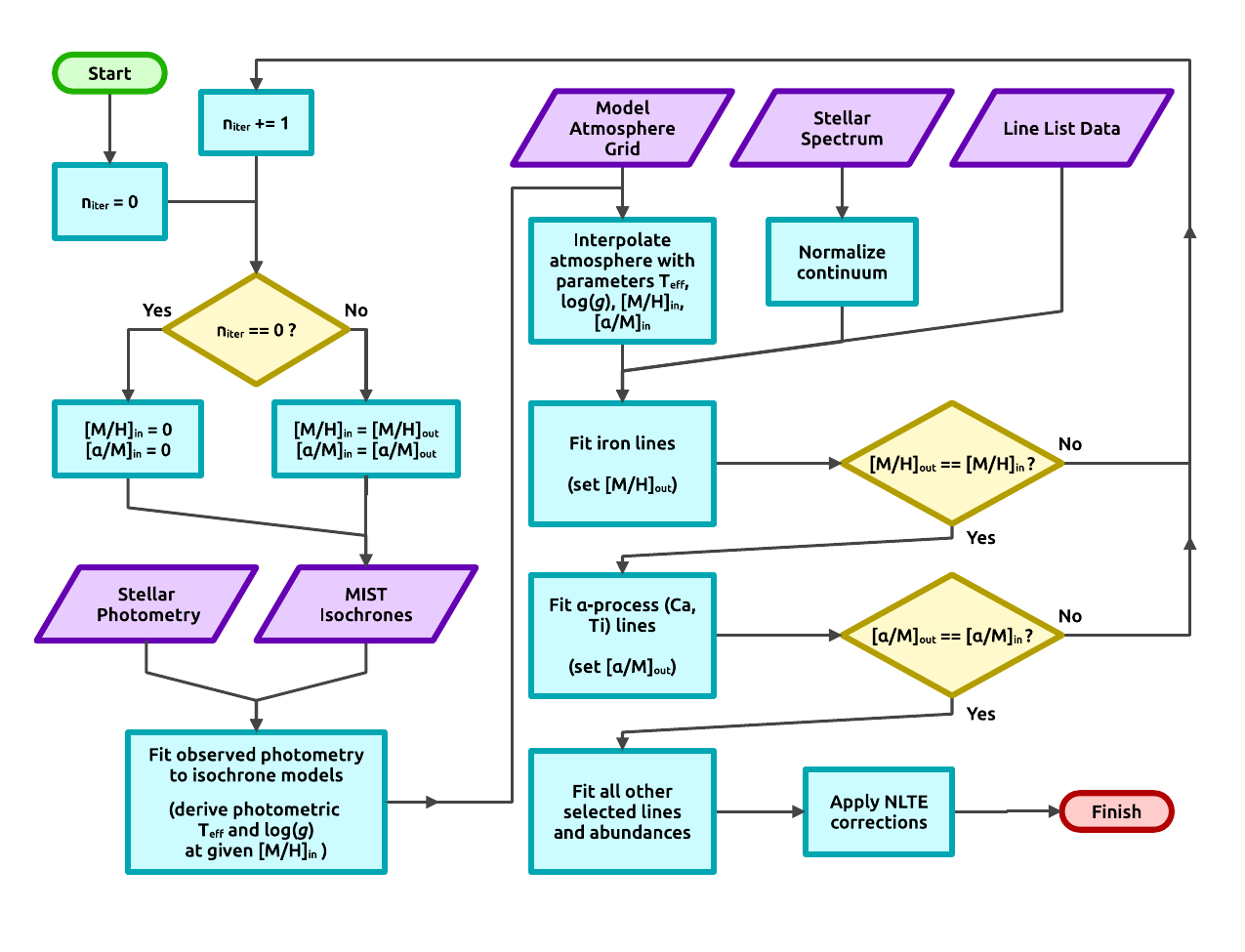}
    \caption{A flow chart of the \mpp algorithm. Blue rectangles represent the execution of a process or function. Yellow diamonds represent a conditional check (e.g. an ``if'' statement). Purple rhomboids represent inputs to the code.}
    \label{fig:FlowChart}
\end{figure*}

\section{The Metal Pipe Algorithm}\label{section:MPP}
In this section we describe the procedure followed by \mpp. A flow chart providing an overview of the procedure is shown in Figure \ref{fig:FlowChart}.


\subsection{Stellar Atmospheric Parameters}\label{alg:params}
To retrieve elemental abundances for a star, \mpp first requires an accurate parametrization of its atmosphere. This is done using the stellar parameters effective temperature and surface gravity (\teff and \logg). Together with bulk metallicity (\monh) and alpha enhancement (\aonm), these parameters are used to interpolate an appropriate model atmosphere from the Phoenix grids included in \mpp. 

Using Astroquery \citep{Ginsburg+2019}, \mpp gathers data from SIMBAD \citep{Wenger+2000} and the Gaia archive. These include UBVRI photometry from various catalogs; J, H, and K photometry from 2MASS \citep{Skrutskie+2006}; and G, Bp, and Rp photometry and parallax from Gaia \citep{Gaia+2016,GaiaDR3}. This photometry is converted from apparent to absolute magnitudes using Gaia parallaxes. Due to the proximity of our sample, we do not invoke any correction models for extinction or interstellar reddening. Once the observed stellar photometry is converted to absolute magnitudes, we compare these magnitudes to synthetic photometry from stellar isochrone models \citep[MIST, ][]{Choi+2016}.

These isochrones are stored as 2-dimensional grids of stellar age and equivalent evolutionary phase (EEP). Each point on the grid represents a model star of a given age and EEP, with associated stellar parameters and synthetic photometry. We compute photometric $\chi^2$ values for each point on this grid according to Equation \ref{eq:chi2}. For a star with $n$ photometric data points, we compare its absolute magnitude $\mathcal{M}_{\rm{b}}^{\rm{obs}}$, with measurement error $\sigma_{\rm{b}}$, to the synthetic photometry $\mathcal{M}_{\rm{b}}^{\rm{model}}$ in each passband $\rm{b}$. This is done for each point $(x,y)$ on the age-EEP grid:

\begin{equation}\label{eq:chi2}
    \chi^2_\nu(x,y) = \frac{1}{n - 2}    \sum_{\rm{b} }\frac{ (\mathcal{M}_{\rm{b}}^{\rm{obs}} - \mathcal{M}_{\rm{b}}^{\rm{model}}(x, y) )^2}{\sigma_{\rm{b}}^2}
\end{equation}

Equations \ref{eq:P} and \ref{eq:norm} are used to convert these $\chi^2_\nu(x,y)$ values into probabilities $P(x,y)$ with normalization constant $A$:

\begin{equation}\label{eq:P}
    P(x,y) = A \exp{(-\frac{1}{2}\chi^2_\nu(x,y) )}
\end{equation}
\begin{equation}\label{eq:norm}
    \frac{1}{A}\sum_{x,y}P(x,y) = 1
\end{equation} 

\begin{figure*}
    \centering
    \includegraphics[width=\linewidth]{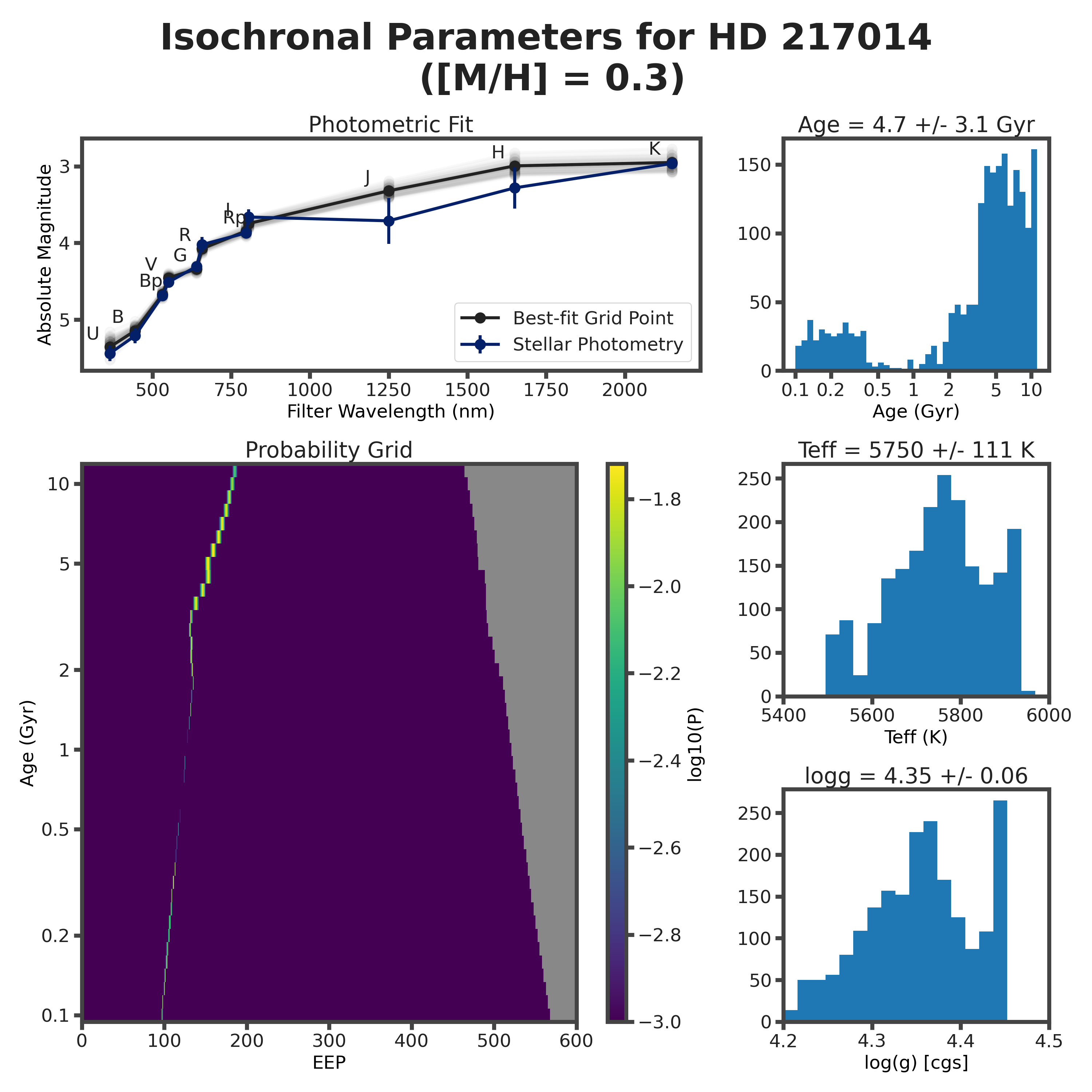}
    \caption{An automatically generated diagnostic plot, illustrating stellar parameter derivation for 51 Peg. \textbf{Bottom left:} Normalized probability grid from $\chi^2_v$ values of the synthetic photometry. \textbf{Top left:} Photometric SED of 51 Peg (``Stellar Photometry''), best-fit synthetic photometry from the isochrone grid (black), and other synthetic photometry values sampled from the probability grid (grey). \textbf{Right:} Histograms of the sampled values for stellar age, effective temperature, and surface gravity.}
    \label{fig:IsoFit}
\end{figure*}

We use rejection sampling of these probabilities to derive stellar parameters. This process involves choosing an age-EEP grid point $(x,y)$ at random, and choosing an associated random number $q \in [0,1]$. If $P(x,y) > q$, stellar parameters are sampled from $(x,y)$. This process repeats until 2,000 samples of each stellar parameter are drawn. We take the median and standard deviation of these samples as the value and error bars of the star's parameters. An illustration of this process is shown in Figure \ref{fig:IsoFit}. Similar figures are automatically generated by \mpp as it analyzes each star.

Since the stellar parameters derived from isochrones are dependent on metallicity, the algorithm initially assumes a solar composition ([M/H] = 0.00, [$\alpha$/M] = 0.00, scaled according to the solar abundances of \citealt{Asplund+2009}), and iterates these values as it progresses (see Section \ref{iterating}).

\subsection{Iteratively Fitting Abundances}
To determine the abundance of an arbitrary element X, we fit synthetic line profiles to the observed spectrum. This is done using a line-by-line, weighted $\chi^2_{\nu}$ minimization of residuals. 

\subsubsection{Fitting A Single Line}

Two free parameters are allowed to vary for each line: the abundance of element X ([X/H]), and a broadening parameter ($v_{\rm{broad}}$). Our use of this singular broadening parameter, which combines the effects of rotational and macroturbulent broadening, is informed by a similar method used by \citet{Brewer+2016}. \texttt{MOOG} handles broadening internally, using two separate broadening kernels. It convolves the synthetic spectrum with a rotational broadening kernel using $v_{\rm{broad}}$ as input by \mpp. It also uses a Gaussian broadening kernel to match the input spectral resolution $R$. $R$ is calculated dynamically for each line based on the spacing of the wavelength grid. We set microturbulence to a constant $v_{\rm{mic}} = 1.5$ km/s. This value of $v_{\rm{mic}}$ was chosen because it most closely reproduced the iron abundance of \citet{Asplund+2009} when \mpp was applied to a solar spectrum (This solar spectrum analysis is described in Section \ref{Subsec:Saas}.) Note that the application of broadening to a spectral feature affects the shape of the line profile and hence goodness of fit, but not the equivalent width.

\mpp will attempt to fit an abundance for each line in the line list for a given element. Several checks are in place to ensure the quality of the fit, and to ignore the results of any lines which are poorly fit. The algorithm by which each line is fit is as follows:

\begin{enumerate}
    \item  \textbf{Set the ``width'' of the line cutout:}  \mpp first synthesizes a spectrum of only the line of interest. With this line as the only source of absorption, it identifies the wavelength region where it produces significant absorption ($\gtrsim 1\%$). It then extends this window by $2\textrm{\AA}$ on either side to capture the continuum. This region is cut out and used to analyze the line.

    \item \textbf{Re-normalize the continuum:} \mpp perform a second-order continuum normalization to account for any imperfections in our earlier normalization process. To do this, it identifies continuum points (where flux $= 1\pm0.05$) in both the observed and synthetic spectrum cutout. It multiplies the observed flux by a constant value which minimizes the $\chi^2_\nu$ of the continuum points. If an insufficient number ($< 4$) of continuum points are identified (e.g. if there are many other nearby spectral lines which obscure the continuum), re-normalization cannot be performed and the line is discarded.
    
    \item \textbf{Cross-correlate the line cutout:} Although the observed spectrum has already been shifted to the stellar rest frame, small variations in the wavelength solution (of order less than 1 pixel), or inaccuracies in line data, can cause a wavelength offset between the observed and synthetic line features. \mpp performs a cross-correlation of the line cutout, re-mapping the observed wavelength data so that line centers in the observed spectrum precisely align with those in the synthetic spectrum.
    
    \item \textbf{Set the $\chi^2$ weight of each data point:} \mpp first generates a synthetic spectrum of the line cutout with a perturbed [X/H] value. By comparing the perturbed and unperturbed spectra, it  generates weights for each data point based on the change in flux at each wavelength as a result of the [X/H] perturbation. The weights are then normalized such that the average weight per point is 1. The result is that \mpp prioritizes regions of the cutout which are sensitive to an abundance change, and ignores those which are minimally affected. This improves the ability of \mpp to handle blended lines and regions with incomplete \texttt{linemake} data.

    \item \textbf{Minimize the $(O-C)$ Residuals:} At the core of \mpp is a single function which takes inputs [X/H] and $v_{\rm{broad}}$, calls MOOG to generate a synthetic spectrum using these values and given stellar parameters, and outputs a weighted $\chi^2_\nu$ value comparing the synthetic line to the observed spectrum. Through numerical minimization of the weighted $\chi^2_\nu$ (defined in Equation \ref{x2eq}), a best-fit value for [X/H] and $v_{\rm{broad}}$ is reached.

    \begin{equation}\label{x2eq}
        \chi^2_\nu = (\frac{1}{n_p-3})\sum_\lambda w(\lambda) \frac
        {(f_{\textrm{obs}}(\lambda) - f_{\textrm{syn}}(\lambda))^2}{(\sigma_{\textrm{obs}}(\lambda)^2 + \sigma_{\textrm{syn}}(\lambda)^2)}
    \end{equation}
    
    where $n_p$ refers to the number of observed data points in the cutout, $w(\lambda)$ is the weight of each point as assigned in step 4, $f(\lambda)$ is the flux at each point, and $\sigma(\lambda)$ is the flux uncertainty at each point. If \mpp is unable to find a minimum $\chi^2_\nu$ value, or the minimum value is too high ($\chi^2_{\nu} > 5$), the line is discarded. 
    
\end{enumerate}

It should be noted that our fit metric is not a ``pure'' $\chi^2_{\nu}$ value. We introduce two additional terms into our calculation: the weight function $w$ and the synthetic spectrum uncertainty $\sigma_{\rm{syn}}$  (see Equations \ref{wl} and \ref{ssyn}). These were added to improve the ability of \mpp to fit deep and/or blended line features.

\begin{equation}\label{wl}
    w(\lambda) \sim \frac{df_{\rm{syn}}}{d \rm{[X/H]}}
\end{equation}

\begin{equation}\label{ssyn}
    \sigma_{\rm{syn}}(\lambda) \sim -\ln{({f_{\rm{syn}}(\lambda)})}
\end{equation}

The weight function encodes sensitivity of a point to changing [X/H]. This encourages \mpp to focus on fitting the most sensitive regions (i.e. line cores) and to ignore badly fit regions it cannot improve by changing [X/H] (e.g. in the case of inaccurate or missing line data for a nearby line from a different species).

The synthetic error bars offset this preference for line cores in deeper lines. $\sigma_{\rm{syn}}(\lambda)$ increases linearly with optical depth (thus with the logarithm of $f(\lambda)$). This discourages \mpp from attempting to over-fit deep line cores (which are often subject to NLTE effects it cannot reproduce). Rather, for deep lines, $\sigma_{\rm{syn}}$ encourages \mpp to focus on fitting the wings accurately, and accept that its LTE synthesis methods will not accurately reproduce deeper line cores.

\begin{figure*}
    \centering
    \includegraphics[width=\linewidth]{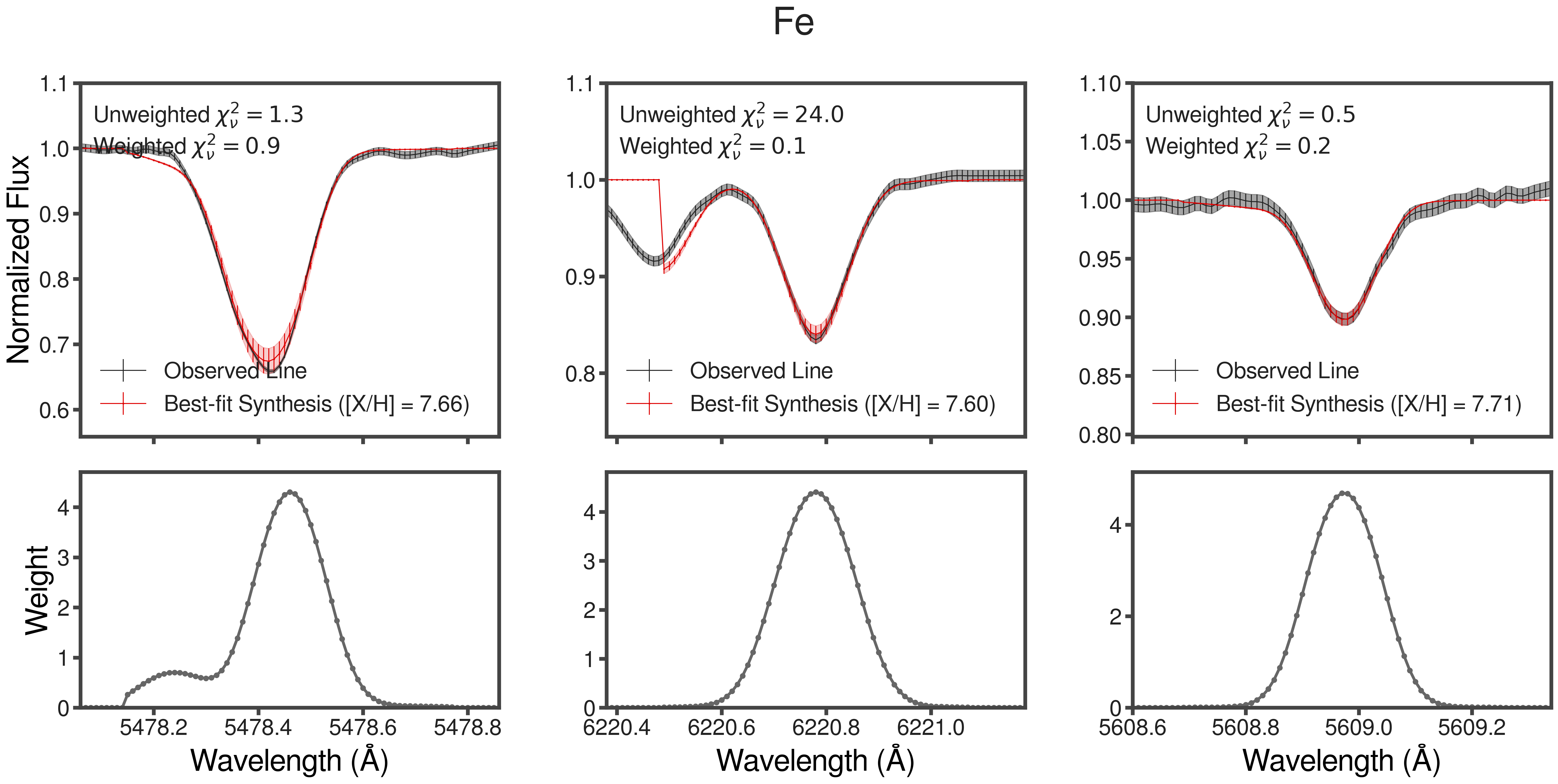}
    \caption{Line fits from \mpp showing the best fit to three Fe lines in 51 Peg. \textbf{Top Row:} The grey line represents the observed line as seen through the spectrograph, and the red line represents the synthetic best fit to the data. The shaded regions around the lines represent $1\sigma$ flux error bars: either intrinsic flux uncertainty in the observed spectrum (grey) or the added uncertainty to the synthetic spectrum as a function of line depth (red). \textbf{Bottom Row:} Weight of each data point in the $\chi^2_\nu$ fit process. Higher weights correspond to regions where the flux is more sensitive to a change in the Fe abundance.}
    \label{fig:LineGallery}
\end{figure*}

As shown in Figure \ref{fig:LineGallery}, \mpp is able to generate synthetic spectral lines which accurately reproduce the observed data. In our analysis of 51 Peg, we were able to fit abundances for 278 iron lines. The resultant median abundance value is [Fe/H] = 7.71 $\pm$ 0.01. Note that while \mpp  uses the uncertainty in the mean to compute errors in the abundance, these errors might be underestimated due to systematic effects.  A more conservative approach is to adopt errors based on the RMS of abundance determinations from different catalogs (See \S\ref{Sec:Verification}).

\begin{figure}
    \centering
    \includegraphics[width=\linewidth]{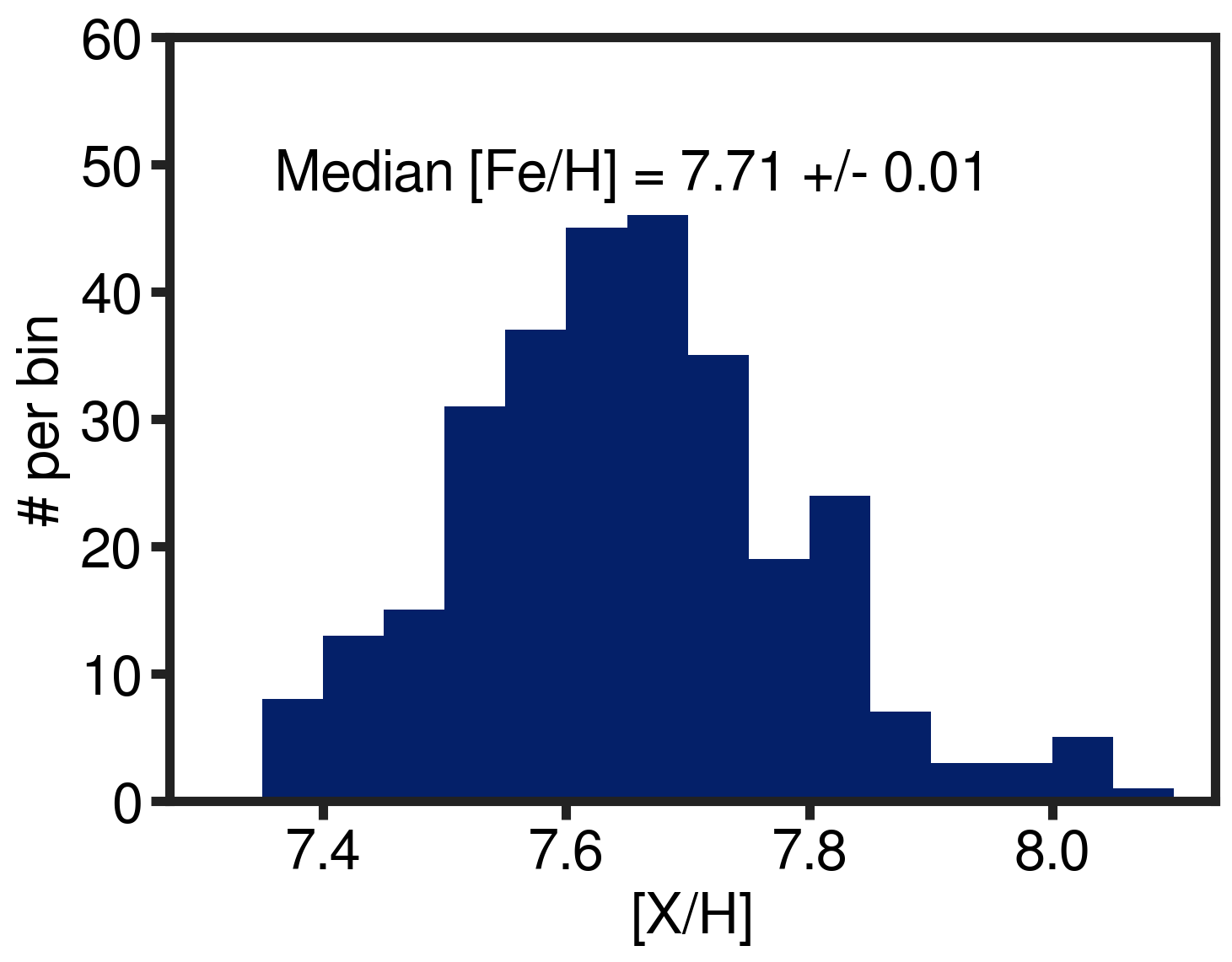}
    \caption{Distribution of abundances derived from Fe lines (n = 278) in 51 Peg.}
    \label{fig:HistAbund}
\end{figure}

\begin{figure}
    \centering
    \includegraphics[width=\linewidth]{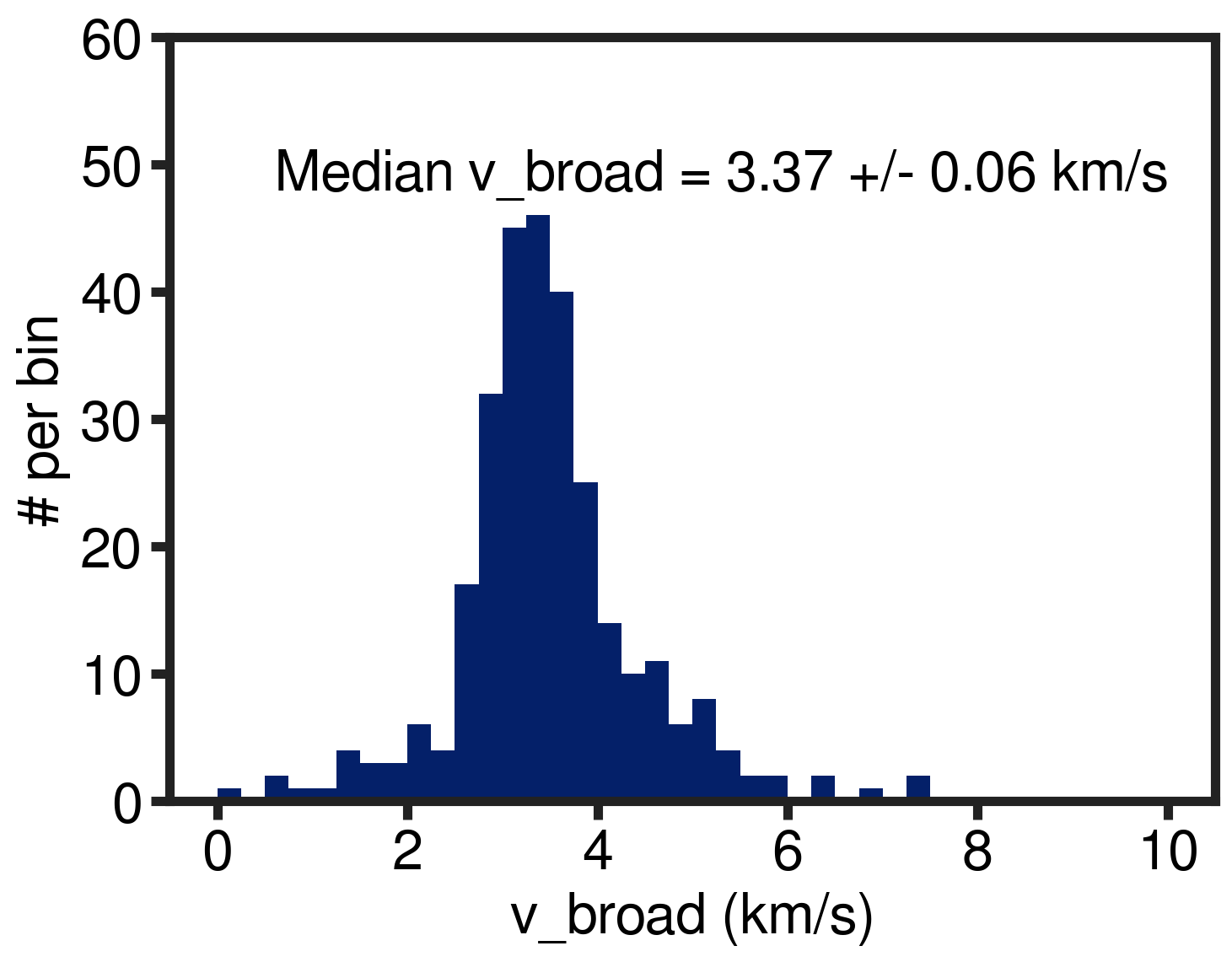}
    \caption{Distribution of $v_{\rm{broad}}$ values for each Fe line fit for 51 Peg.}
    \label{fig:HistChi2}
\end{figure}

A histogram of the abundance values for each Fe line fit by \mpp in 51 Peg is shown in Figure \ref{fig:HistAbund}. A histogram of $v_{\rm{broad}}$ values for these same lines is shown in Figure \ref{fig:HistChi2}.

\subsubsection{Iterating [M/H] and [$\alpha$/M]}\label{iterating}

We assume the stellar iron abundance is equivalent to its bulk metallicity (i.e. [Fe/H] = [M/H]). To fit [M/H], \mpp fits abundances for every iron line in its line list. We take the median abundance value from these lines as the star's ``output'' [M/H], with the uncertainty of the mean as its uncertainty. If this output [M/H] does not fall within $1\sigma$ of the ``input'' [M/H] (i.e. the [M/H] associated with the isochronal stellar parameters and model atmosphere), we set the input [M/H] of the next iteration equal to the output [M/H] of the previous iteration. This is repeated until the input and output values are consistent with each other (see Figure \ref{fig:FlowChart}.)

Once the algorithm converges [M/H], we perform a similar procedure to converge the star's alpha enhancement. Alpha enhancement in \mpp is defined as the average abundance of calcium and titanium with respect to iron (i.e. [$\alpha$/M] = ([Ca/Fe] + [Ti/Fe])/2). We allow the input and output [$\alpha$/M] values to iterate until they are consistent to within $1\sigma$. Note that the opacity effects of the alpha elements can impact the [M/H] value derived from Fe lines. We account for this by allowing [M/H] to continue to vary while [$\alpha$/M] converges.

\subsubsection{All Other Abundances}

In addition to Fe, Ca, and Ti, we consider abundances of C, O, Na, Mg, Al, and Si in this work (although N and P are important elements in the CHNOPS family, they are particularly challenging to measure, and we ignore them for this work.) We ignore any impact of these elements on opacities in the stellar atmosphere because our chosen model atmosphere grid does not have models with varied abundances of these elements. Therefore, once [M/H] and [$\alpha$/M] have converged, we perform a single fit to each line of each species, taking the median abundance value for all lines of a given species as its corresponding [X/H].  Note that \mpp generates figures of the distribution of best-fit values for each line in each element, and also the distribution of weighted $\chi^2$ values, similar to Figure \ref{fig:HistAbund}, so that the reported abundance values can be inspected for quality and consistency.

\subsubsection{NLTE Corrections}\label{section:nlte}

While non-LTE effects on line formation are negligible for some elements (e.g. Fe, Si), some of our abundance values are significantly affected by the assumption of LTE. Thus, once the analysis is complete, we apply abundance corrections for non-LTE effects to oxygen \citep{Amarsi+2019}, calcium \citep{Mashonkina+2007}, and titanium \citep{Bergemann2011}. In each case, these corrections served to improve the agreement of our abundances with those in the literature.

\section{Benchmarking Metal Pipe against Literature Sources}\label{Sec:Verification}

To test the accuracy of \mpp, we analyzed a subset of the HIRES spectra which were used to create the 2016 version of the SPOCS catalog \citep{Brewer+2016}. In total, our data set contains spectra from 503 stars, of which 104 are known to host planets. Table \ref{table:data} shows our final reported values for each star. In the following section, we compare our results to those of \citet{Brewer+2016} and \citet{Adibekyan+2012}.




\begin{splitdeluxetable*}{lrrrrrrrrrrBrrrrrrrrrrBrrrrrrrrrr}
\tablecaption{
    Final Stellar Abundance Table \label{table:data}
}
\tablehead{
    \colhead{Name} &
    \colhead{ \teff} &
    \colhead{$\sigma(T_{\rm eff})$} &
    \colhead{\logg} &
    \colhead{$\sigma(\log{(g)})$} &
    \colhead{$M/M_{\odot}$} &
    \colhead{$\sigma(M/M_{\odot})$} &
    \colhead{$R/R_{\odot}$} &
    \colhead{$\sigma(R/R_{\odot})$} &
    \colhead{$L/L_{\odot}$} &
    \colhead{$\sigma(L/L_{\odot})$} &
    \colhead{$\textrm{[Fe/H]}$} &
    \colhead{$\sigma(\textrm{[Fe/H]})$} &
    \colhead{$\textrm{[Ca/H]}$} &
    \colhead{$\sigma(\textrm{[Ca/H]})$} &
    \colhead{$\textrm{[Ti/H]}$} &
    \colhead{$\sigma(\textrm{[Ti/H]})$} &
        \colhead{$\textrm{[Mg/H]}$} &
    \colhead{$\sigma(\textrm{[Mg/H]})$} &
        \colhead{$\textrm{[Si/H]}$} &
    \colhead{$\sigma(\textrm{[Si/H]})$} &
        \colhead{$\textrm{[C/H]}$} &
    \colhead{$\sigma(\textrm{[C/H]})$} &
        \colhead{$\textrm{[O/H]}$} &
    \colhead{$\sigma(\textrm{[O/H]})$} &
        \colhead{$\textrm{[S/H]}$} &
    \colhead{$\sigma(\textrm{[S/H]})$} &
        \colhead{$\textrm{[Na/H]}$} &
    \colhead{$\sigma(\textrm{[Na/H]})$} &
        \colhead{$\textrm{[Al/H]}$} &
    \colhead{$\sigma(\textrm{[Al/H]})$}
    }

\startdata
        HD 100180 & 6018 & 101 & 4.39 & 0.05 & 1.07 & 0.05 & 1.10 & 0.04 & 1.42 & 0.03 & 0.02 & 0.01 & 0.03 & 0.03 & 0.00 & 0.02 & -0.06 & 0.04 & -0.01 & 0.04 & 0.01 & 0.03 & -0.02 & 0.06 & 0.00 & 0.01 & -0.05 & 0.08 & -0.04 & 0.09 \\ 
        HD 100623 & 5258 & 30 & 4.61 & 0.02 & 0.78 & 0.02 & 0.72 & 0.01 & 0.36 & 0.01 & -0.33 & 0.01 & -0.27 & 0.04 & -0.21 & 0.02 & -0.21 & 0.04 & -0.32 & 0.03 & -0.39 & 0.05 & -0.30 & 0.01 & -0.35 & 0.01 & -0.27 & 0.04 & -0.19 & 0.07 \\ 
        HD 10145 & 5703 & 75 & 4.39 & 0.04 & 0.95 & 0.04 & 1.02 & 0.03 & 1.00 & 0.02 & 0.03 & 0.01 & 0.01 & 0.04 & 0.10 & 0.02 & 0.07 & 0.04 & 0.01 & 0.03 & -0.02 & 0.04 & -0.01 & 0.04 & -0.06 & 0.01 & -0.07 & 0.06 & 0.11 & 0.06 \\ 
        HD 101472 & 6142 & 98 & 4.34 & 0.05 & 1.03 & 0.04 & 1.14 & 0.04 & 1.66 & 0.02 & -0.14 & 0.01 & -0.06 & 0.03 & -0.15 & 0.03 & -0.23 & 0.04 & -0.15 & 0.04 & -0.10 & 0.03 & 0.08 & 0.05 & -0.10 & 0.01 & -0.17 & 0.10 & -0.30 & 0.09 \\ 
        HD 101501 & 5511 & 56 & 4.53 & 0.04 & 0.91 & 0.03 & 0.86 & 0.02 & 0.61 & 0.02 & -0.02 & 0.01 & -0.05 & 0.04 & -0.06 & 0.02 & -0.09 & 0.04 & -0.04 & 0.03 & -0.04 & 0.02 & 0.13 & 0.06 & -0.07 & 0.01 & -0.13 & 0.06 & -0.10 & 0.09 \\ 
\enddata
\tablecomments{The full catalog is available in machine-readable format.}

\end{splitdeluxetable*}

\subsection{Setting a Solar Reference Point}\label{Subsec:Saas}
The solar abundance values of \citet{Asplund+2009} are used by \texttt{MOOG}, the PHOENIX atmosphere grid, and MIST isochrone grid that \mpp uses. However, these abundances are not necessarily equal to the abundances that \mpp would report if run on an input solar spectrum. To ensure that our solar abundances are self-consistent with the rest of the abundances in our catalog, we used \mpp to analyze a HIRES solar spectrum captured using reflected light from the asteroid Vesta. 

Using the nominal solar values for \teff and \logg (5777K and 4.44 respectively), we allowed \mpp to analyze the solar spectrum just as it would for any other star. Table \ref{table:solar} shows the results of this ``Sun as a Star'' (SaaS) analysis. With this result, we can more accurately set a solar reference point for our abundances. To remove systematic offsets, we subtract the differences between the \mpp-determined and \citet{Asplund+2009} solar abundances from the rest of our catalog before comparing our abundances with the literature. The largest systematic offset was for Al (0.16 dex).

Note that the solar abundances listed in Table \ref{table:solar} are applicable only to HIRES spectra analyzed with \mpp, analyzed with identical line lists to those current at the time of writing. We plan to include similar SaaS analyses in future works with other spectrographs. Note that this SaaS analysis effectively removes systematic abundance errors in our results for stars that are like the Sun, but it may be less reliable for stars with \teff and \logg values that differ significantly from the Sun.

\begin{table}[t]
\footnotesize
    \centering
    \begin{tabular}{|c|c|c|}
    \hline
    \textbf{Element} & \multicolumn{2}{c|}{\textbf{Solar Abundances}} \\
       & \textbf{A09} & \textbf{MPP SaaS} \\
      \hline
      C  &  8.43 &  8.40  \\
      O  &  8.69 &  8.72  \\
      Na &  6.24 &  6.29  \\
      Mg &  7.60 &  7.52  \\
      Al &  6.45 &  6.29  \\
      Si &  7.51 &   7.55  \\
      Ca &  6.34 &   6.34  \\
      Ti &  4.95 &   4.98  \\
      Fe &  7.50 &   7.49  \\
      \hline
    \end{tabular}
    \caption{Solar abundances reported by \citet{Asplund+2009}, and abundance values derived for the Sun by \mpp.}
    \label{table:solar}
\end{table}

\subsection{Stellar Parameters}
\begin{figure}
    \centering
    \includegraphics[width=\linewidth]{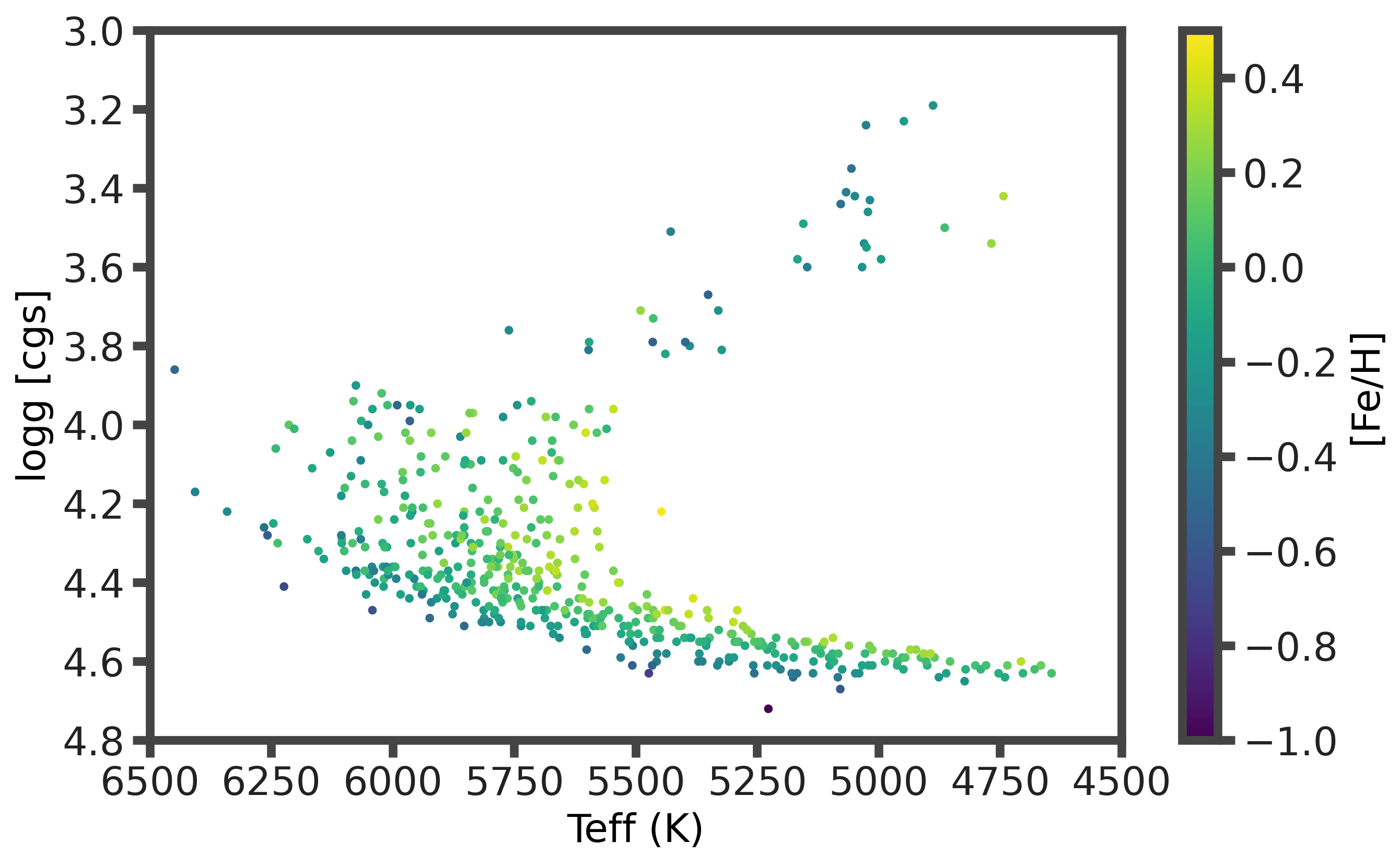}
    \caption{Effective temperatures, surface gravities, and iron abundances determined from \mpp.}
    \label{fig:HR}
\end{figure}

Our values for \teff and \logg were derived using public data from large photometric surveys (see Section \ref{alg:params}). A plot of \logg vs. \teff, with color indicating [Fe/H], is shown in Figure \ref{fig:HR}.  This is a different method from the one used for the SPOCS catalog, where \citet{Brewer+2016} used depth and pressure broadening profiles of strong spectral lines (e.g. H-$\alpha$, H-$\beta$). \citet{Adibekyan+2012} used yet another method: tuning the value of \teff until the abundances from iron lines show no trend with excitation potential or equivalent width, and tuning the value of \logg until the abundances derived from Fe I and Fe II lines are equivalent.

\subsubsection{Results Comparison}

\begin{figure*}
    \centering
    \includegraphics[width=\linewidth]{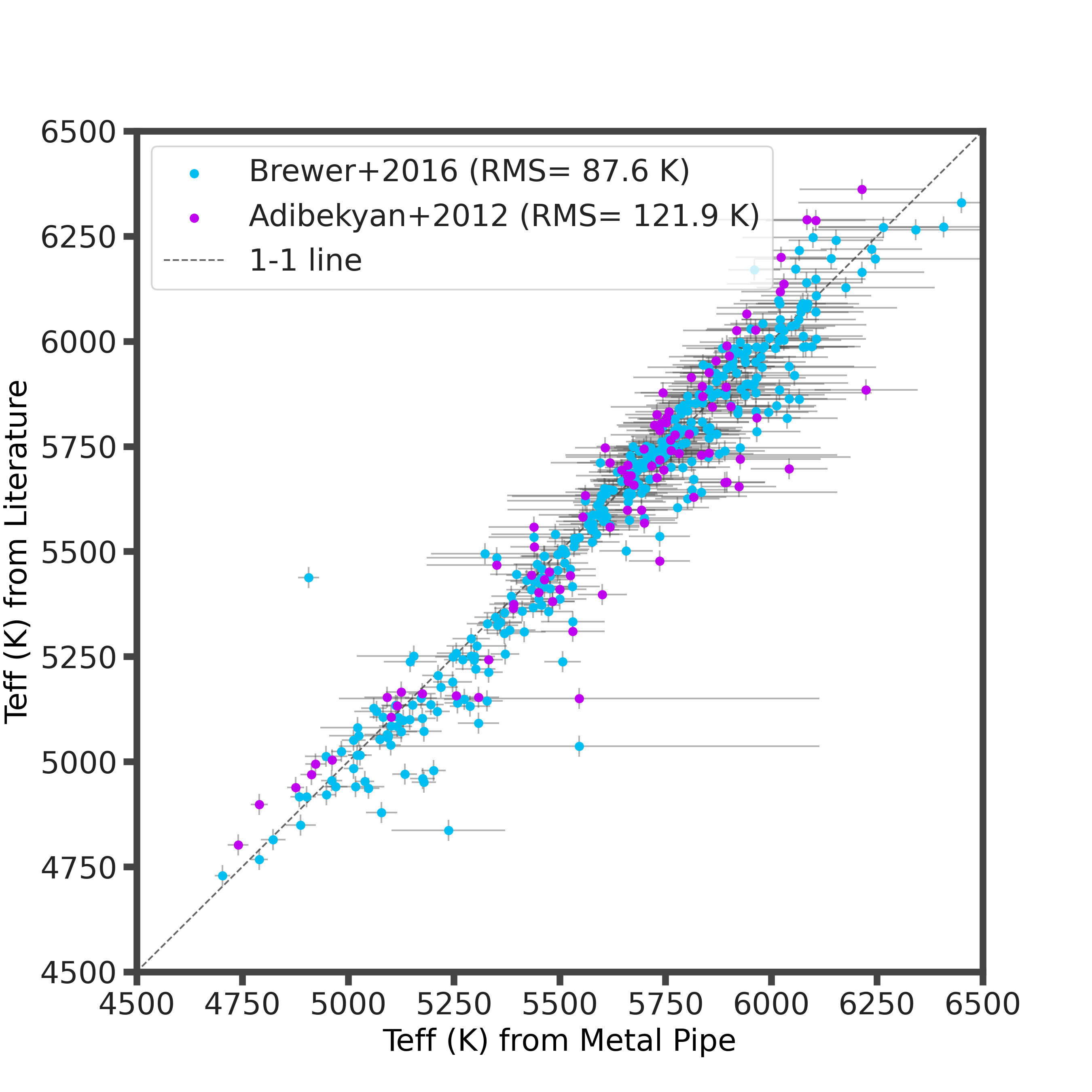}
    \caption{Comparison of effective temperatures derived by \mpp and by literature sources}
    \label{fig:teff}
\end{figure*}

\begin{figure*}
    \centering
    \includegraphics[width=\linewidth]{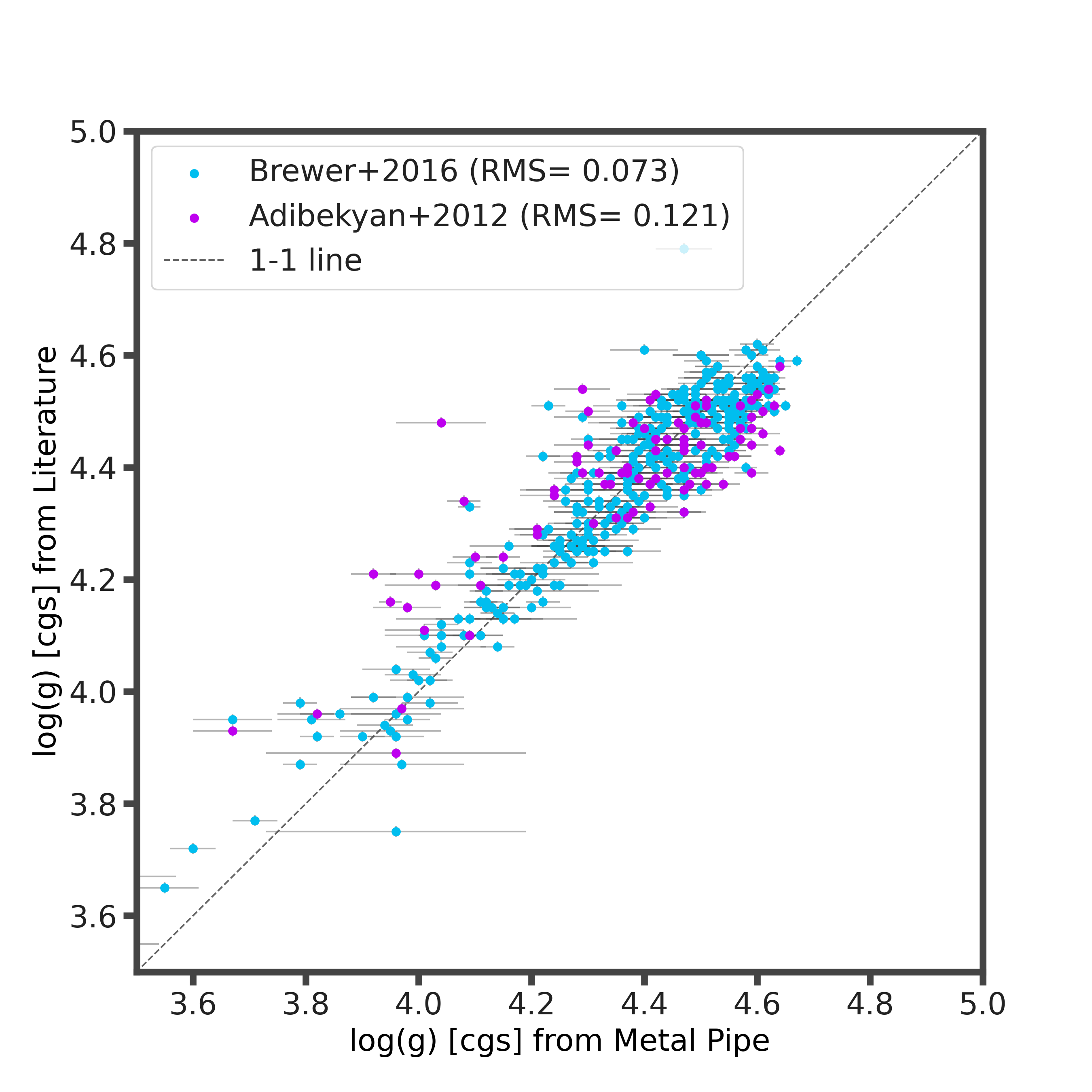}
    \caption{Comparison of surface gravities derived by \mpp and by literature sources}
    \label{fig:logg}
\end{figure*}

Our values for \teff and \logg are largely consistent with those of \citet{Brewer+2016} and \citet{Adibekyan+2012}, as can be seen in Figure \ref{fig:teff}, Figure \ref{fig:logg}, and Table \ref{table:teffComp}. We find that the scatter between our work and literature catalogs is comparable to that found by \citet{Petigura+2017} between their work and other literature catalogs. 

In the absence of any systematics, we expect the RMS scatter between different methods to be $\sim100$ K for \teff, and $\sim0.05$ dex for \logg \citep{Tayar+2022}. While our \teff values have an RMS scatter with the literature which is comparable to this lower limit, the RMS scatter of our \logg values is slightly higher than expected from uncertainties on stellar mass and radius alone. This may be attributed to our use of strictly photometric observations to derive stellar surface gravities. The extra computational steps involved in deriving log(g) from distance and photometry propagate errors which, both increases our derived uncertianties in log(g) and the RMS scatter compared to  literature \logg values.  More direct methods of probing log(g) (e.g. interferometry and asteroseismology) typically reduce the intrinsic uncertainty in \logg, but these techniques are only applicable to a subset of stars.  The width of absorption lines is also affected by \logg, although the measure of surface gravity can be easily confused with other line-broadening mechanisms (e.g., rotation, macroturbulence, etc.)  A possible future approach would be to iteratively fit the photometry and high-resolution spectrum when determining stellar physical parameters, as this might reduce the intrinsic errors and the scatter in \logg.

\begin{table}[t]
    \footnotesize
    \centering
    \begin{tabular}{|c|rr|rr|}
    \hline
      & \multicolumn{2}{c|}{\textbf{This Work - B16}} & \multicolumn{2}{c|}{\textbf{This Work - A12}} \\
      \hline
      \textbf{Element} & \textbf{Median} & \textbf{Std. Dev.} & \textbf{Median} & \textbf{Std. Dev.} \\
      \hline
      \teff (K) &  23 & $\pm$88  &   11  &  $\pm$122  \\
      \logg [cgs] & 0.00 & $\pm$0.07 &   -0.01  &  $\pm$0.12  \\
      \hline
    \end{tabular}
    \caption{Summary of our \teff and \logg distributions compared to the literature values, for stars common between this work and literature catalogs (either B16 \citep{Brewer+2016} or A12 \citep{Adibekyan+2012}).}
    \label{table:teffComp}
\end{table}

\subsubsection{Stellar Parameter Outliers}
For a small number of stars, our \teff and \logg values differed significantly from the literature values. Stars that suffer from this issue generally fall into one of two categories:

\begin{itemize}
    \item \textbf{Young stars:} The gas and dust around a young star absorb some of the star's emitted light. Because shorter wavelengths of light are more readily absorbed, this makes the star appear ``redder'' than it actually is. Additionally, if this material is hot enough, it can produce its own significant infrared radiation. These phenomena affect the ability of \mpp to fit the stellar photometry, and may prevent it from converging on an abundance solution. We have noted these effects in the following stars, which we removed from our final reported catalog due to non-convergence: HD 105 \citep[see e.g.][]{Hollenbach+2005}, and HD 143006 \citep[see e.g.][]{Benisty+2018}.

    \item \textbf{Binary Contamination:} If a star is part of a multiple-star system with low angular separation, the light of its companion star(s) may contaminate the photometry of the intended target. In our sample, HD 21847 likely suffered photometric contamination from a nearby companion\footnote{SIMBAD lists this object as a double star} and was removed from our catalog due to non-convergence. 
\end{itemize}

\subsection{Abundances}

Assuming identical stellar parameters, discrepancies in abundance measurements arise largely as a result of variations in the data and models which make up the analysis procedure \citep[e.g.][]{Hinkel+2016,BC2019,Jofre+2018}. Each of these variations (summarized in Table \ref{table:models}) introduces its own idiosyncratic differences into the final abundances which are reported by each study.

\subsubsection{Results Comparison}
\begin{figure*}
    \centering
    \includegraphics[width=\linewidth]{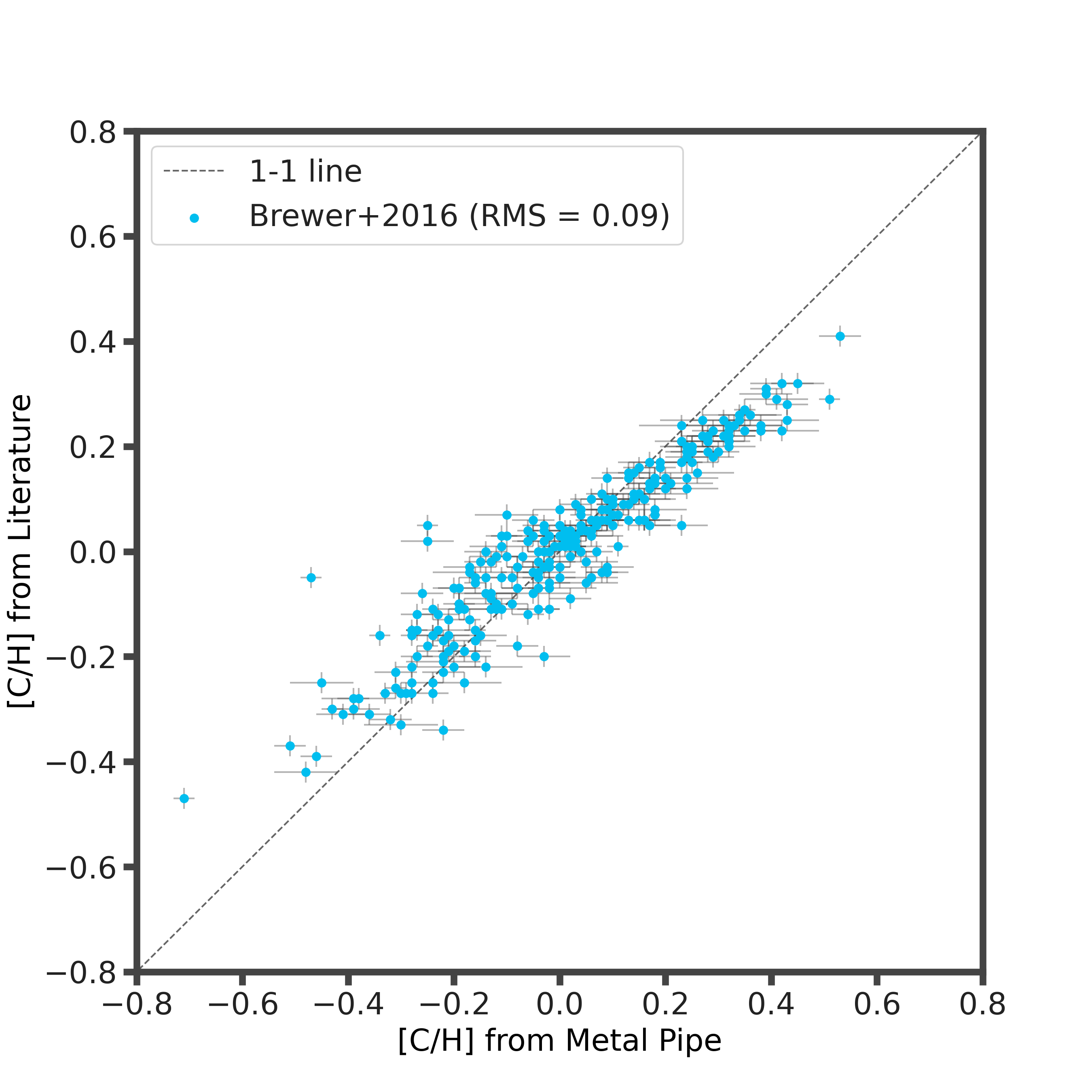}
    \caption{Comparison of carbon abundances derived by \mpp and by literature sources}
    \label{fig:6}
\end{figure*}

\begin{figure*}
    \centering
    \includegraphics[width=\linewidth]{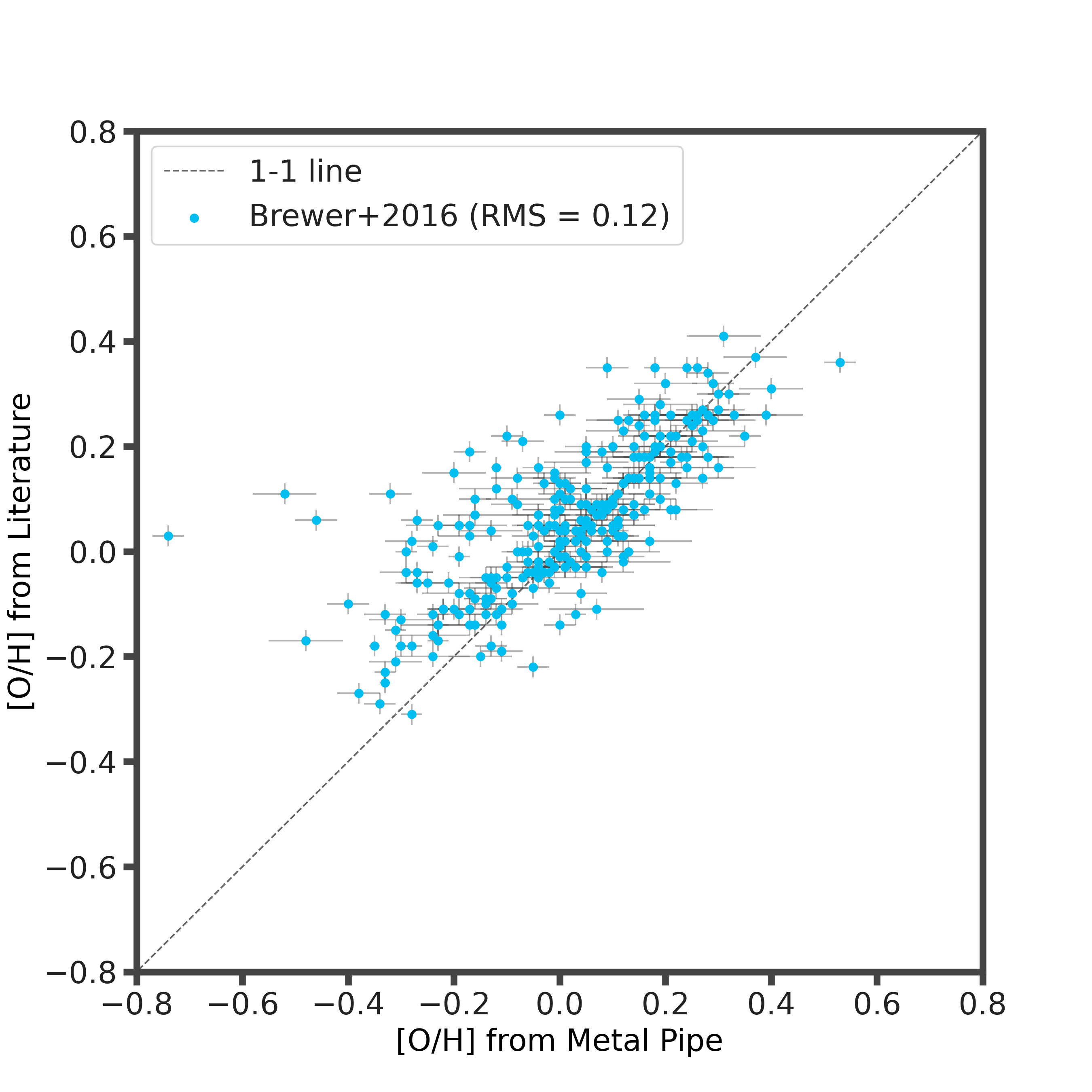}
    \caption{Comparison of oxygen abundances derived by \mpp and by literature sources}
    \label{fig:8}
\end{figure*}

\begin{figure*}
    \centering
    \includegraphics[width=\linewidth]{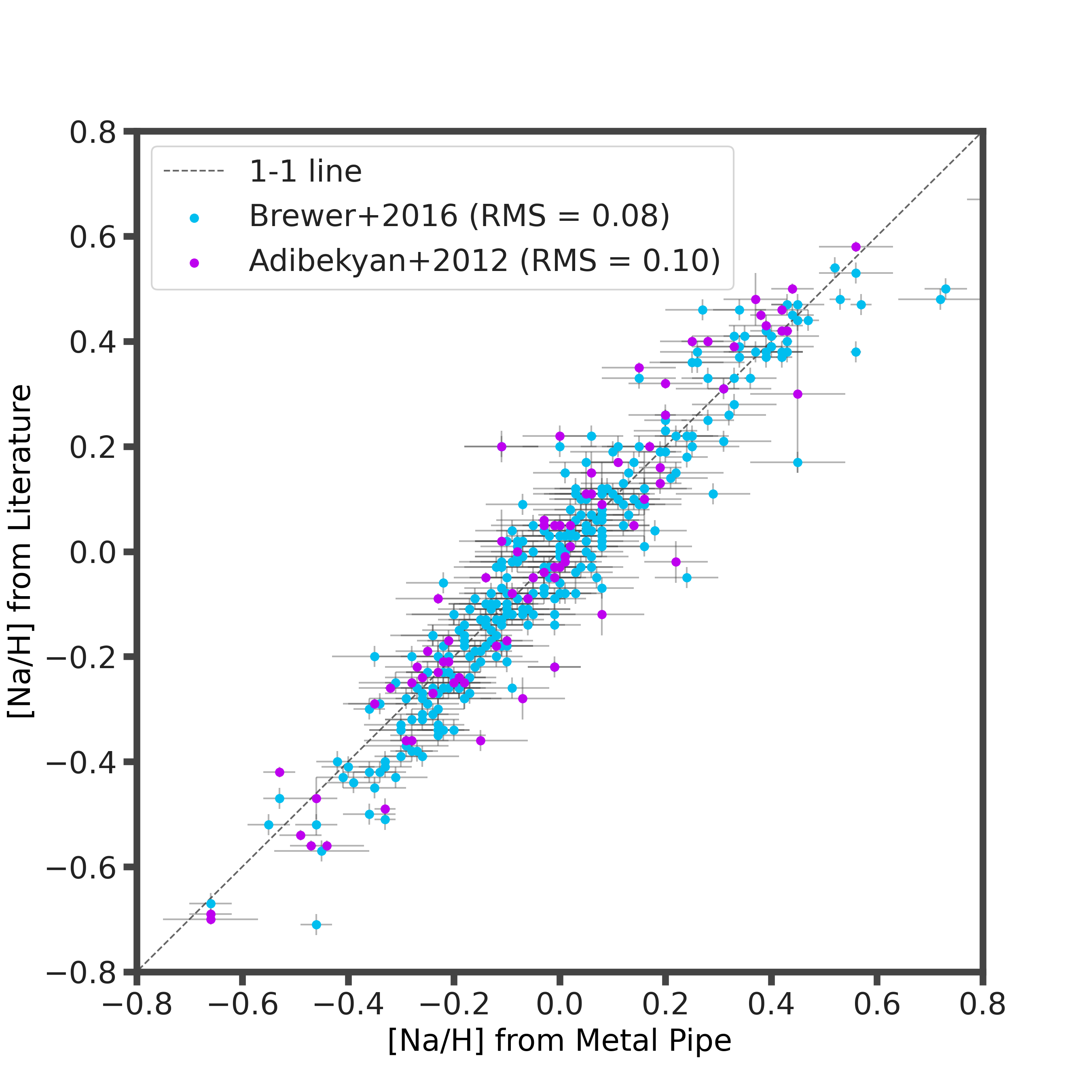}
    \caption{Comparison of sodium abundances derived by \mpp and by literature sources}
    \label{fig:11}
\end{figure*}

\begin{figure*}
    \centering
    \includegraphics[width=\linewidth]{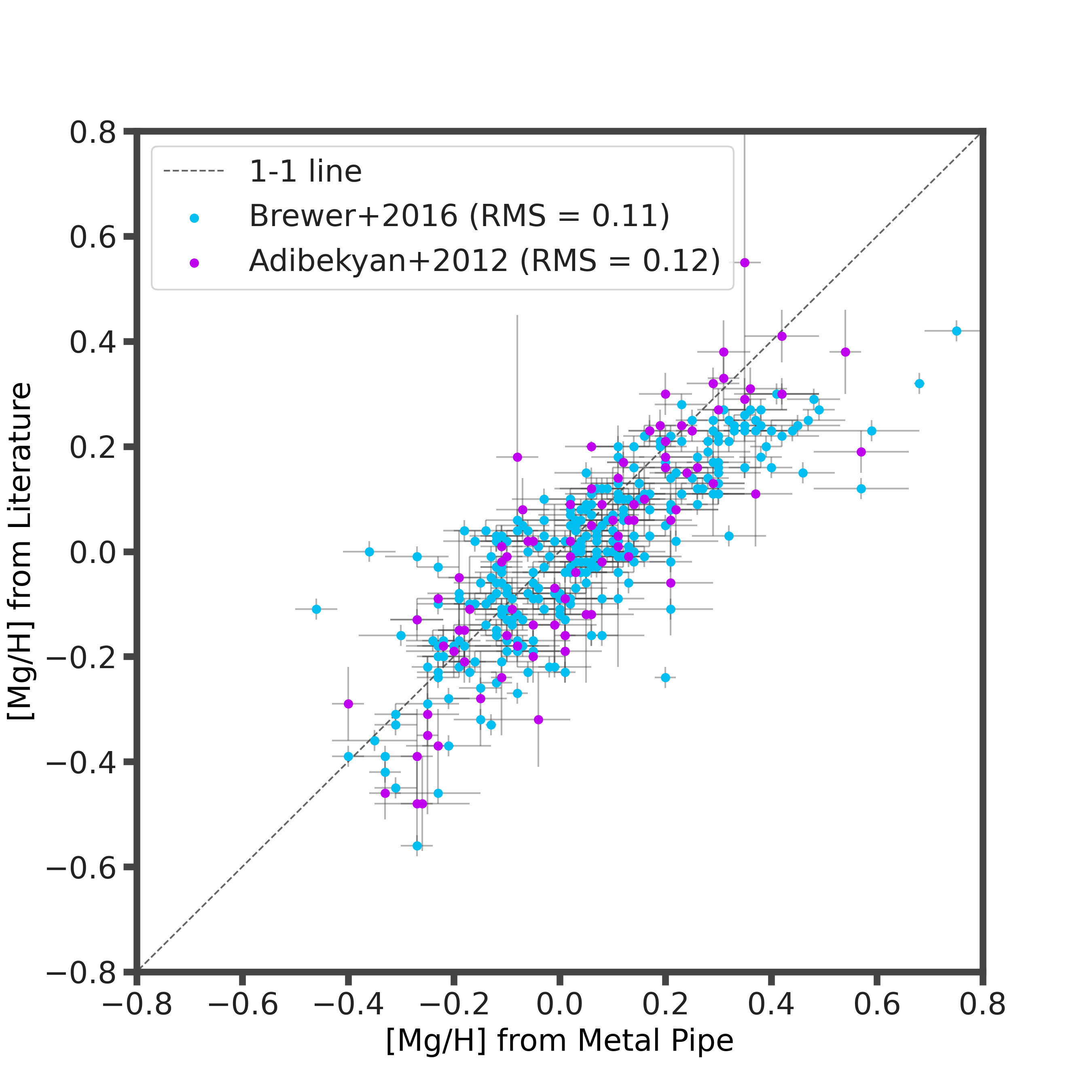}
    \caption{Comparison of magnesium abundances derived by \mpp and by literature sources}
    \label{fig:12}
\end{figure*}

\begin{figure*}
    \centering
    \includegraphics[width=\linewidth]{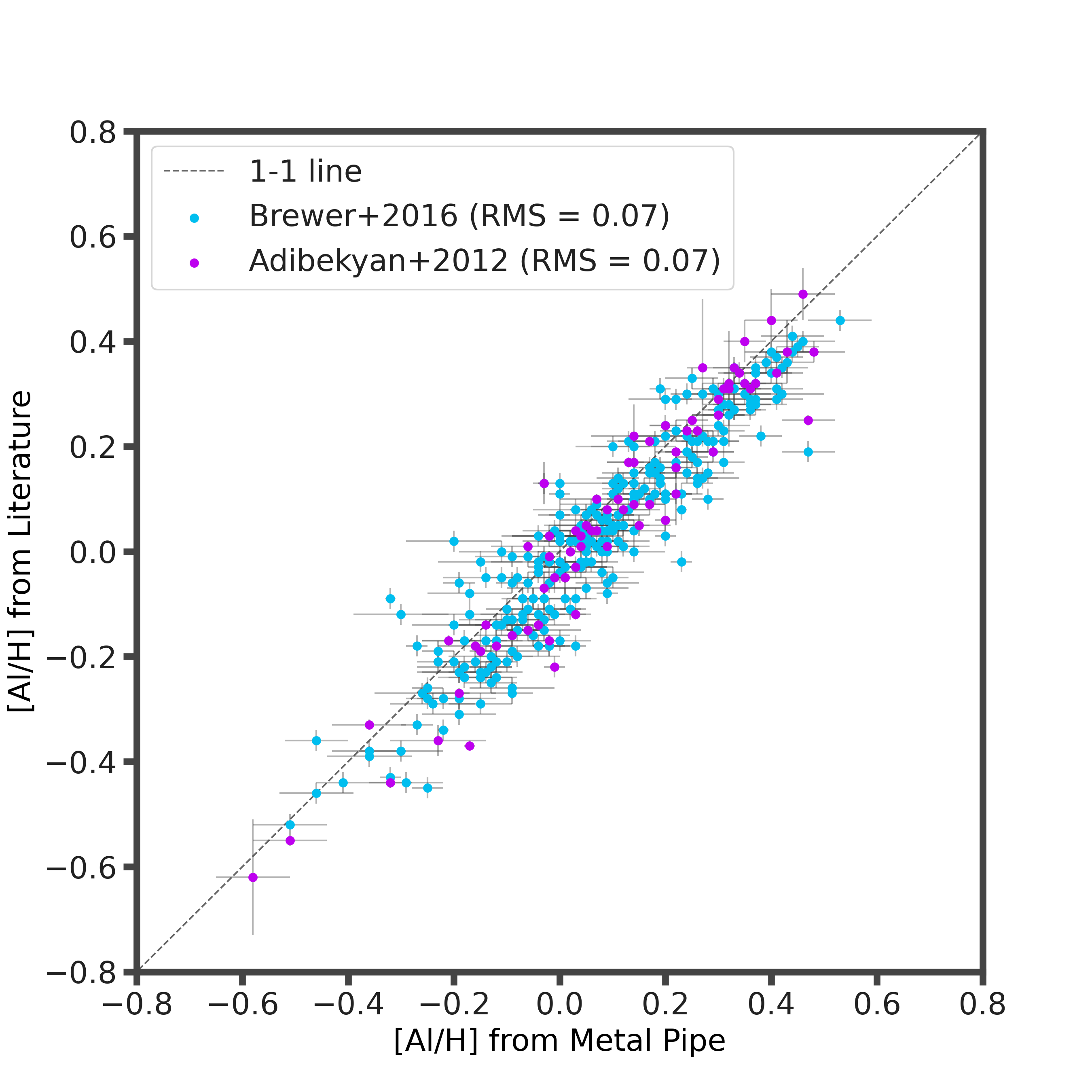}
    \caption{Comparison of aluminum abundances derived by \mpp and by literature sources}
    \label{fig:13}
\end{figure*}

\begin{figure*}
    \centering
    \includegraphics[width=\linewidth]{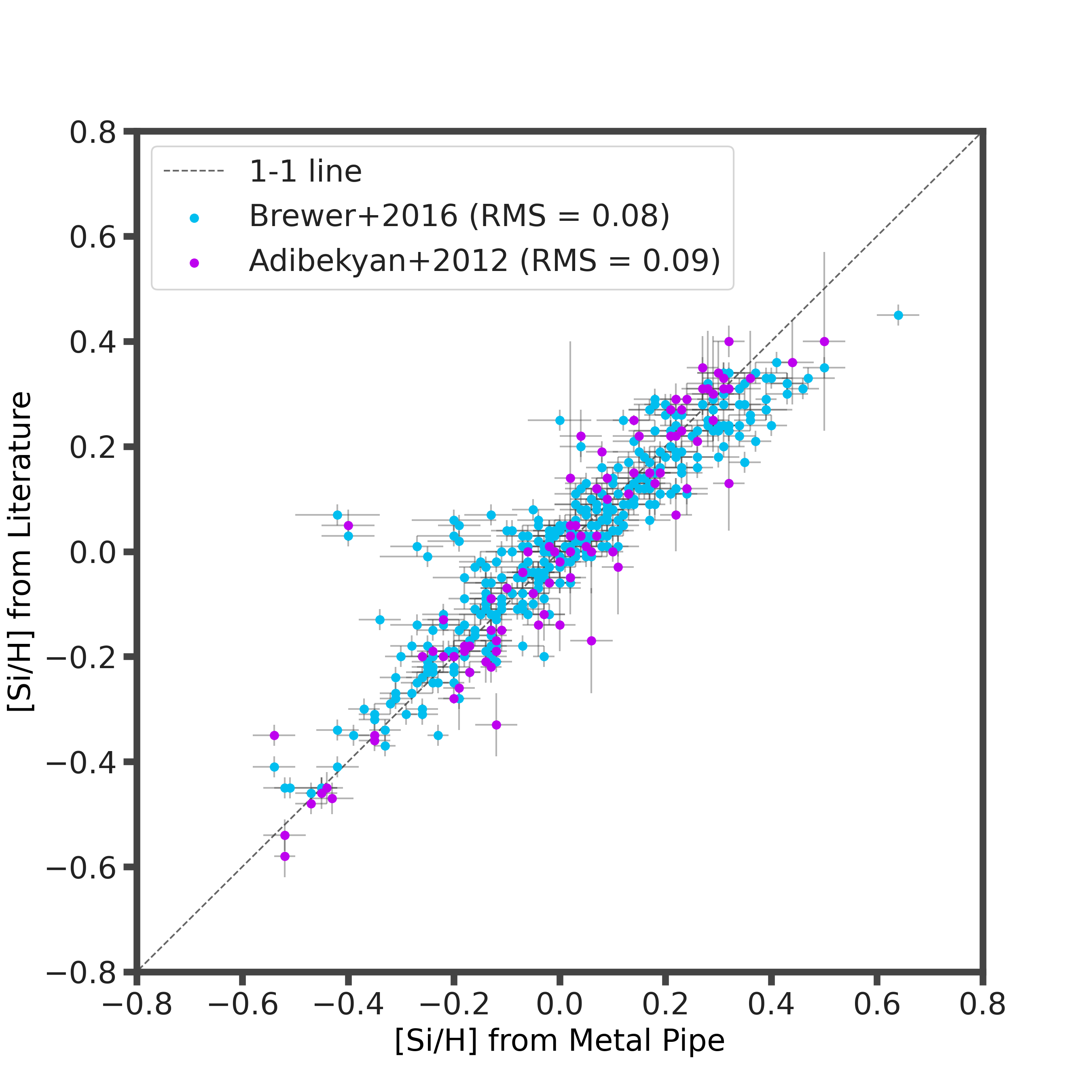}
    \caption{Comparison of silicon abundances derived by \mpp and by literature sources}
    \label{fig:14}
\end{figure*}

\begin{figure*}
    \centering
    \includegraphics[width=\linewidth]{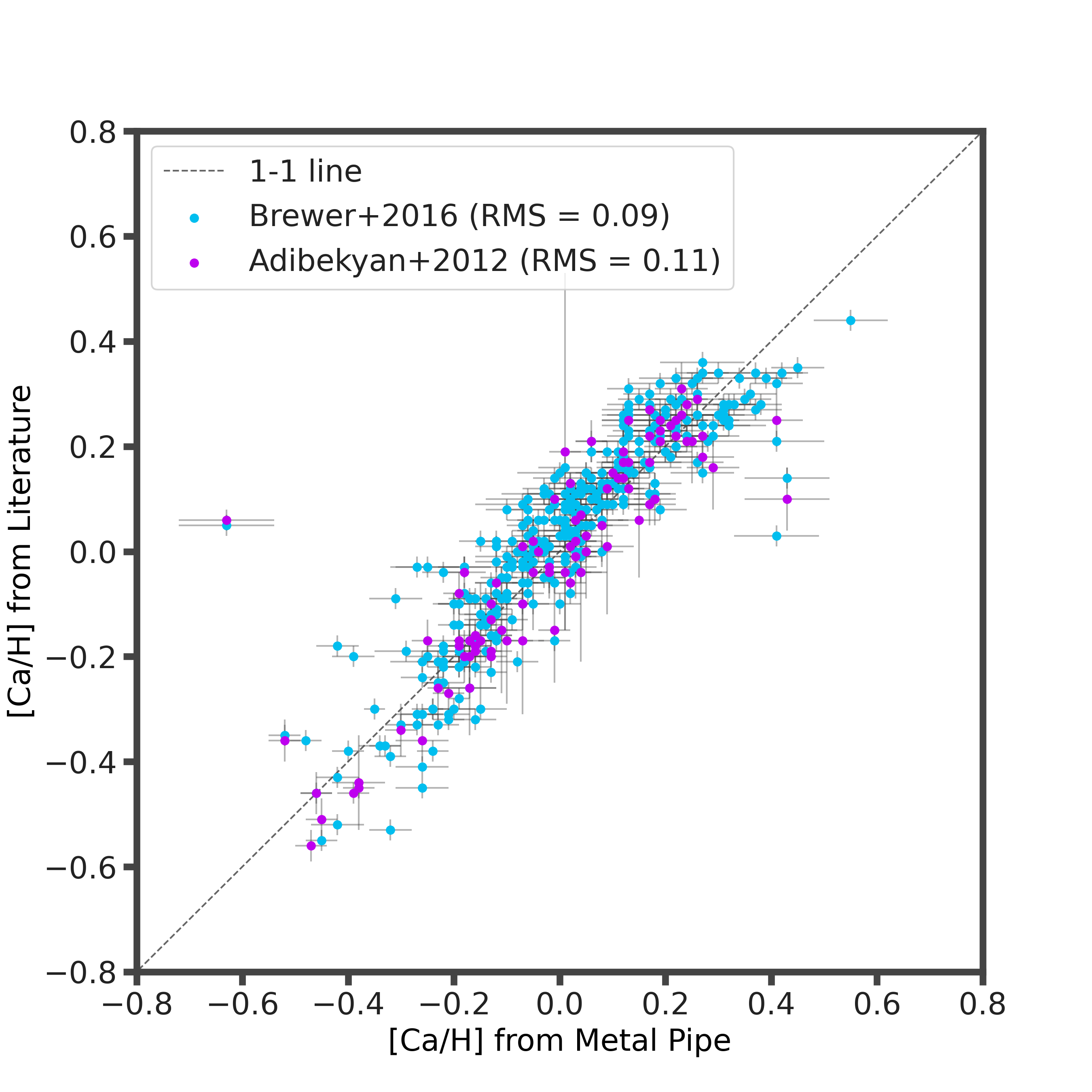}
    \caption{Comparison of calcium abundances derived by \mpp and by literature sources}
    \label{fig:20}
\end{figure*}

\begin{figure*}
    \centering
    \includegraphics[width=\linewidth]{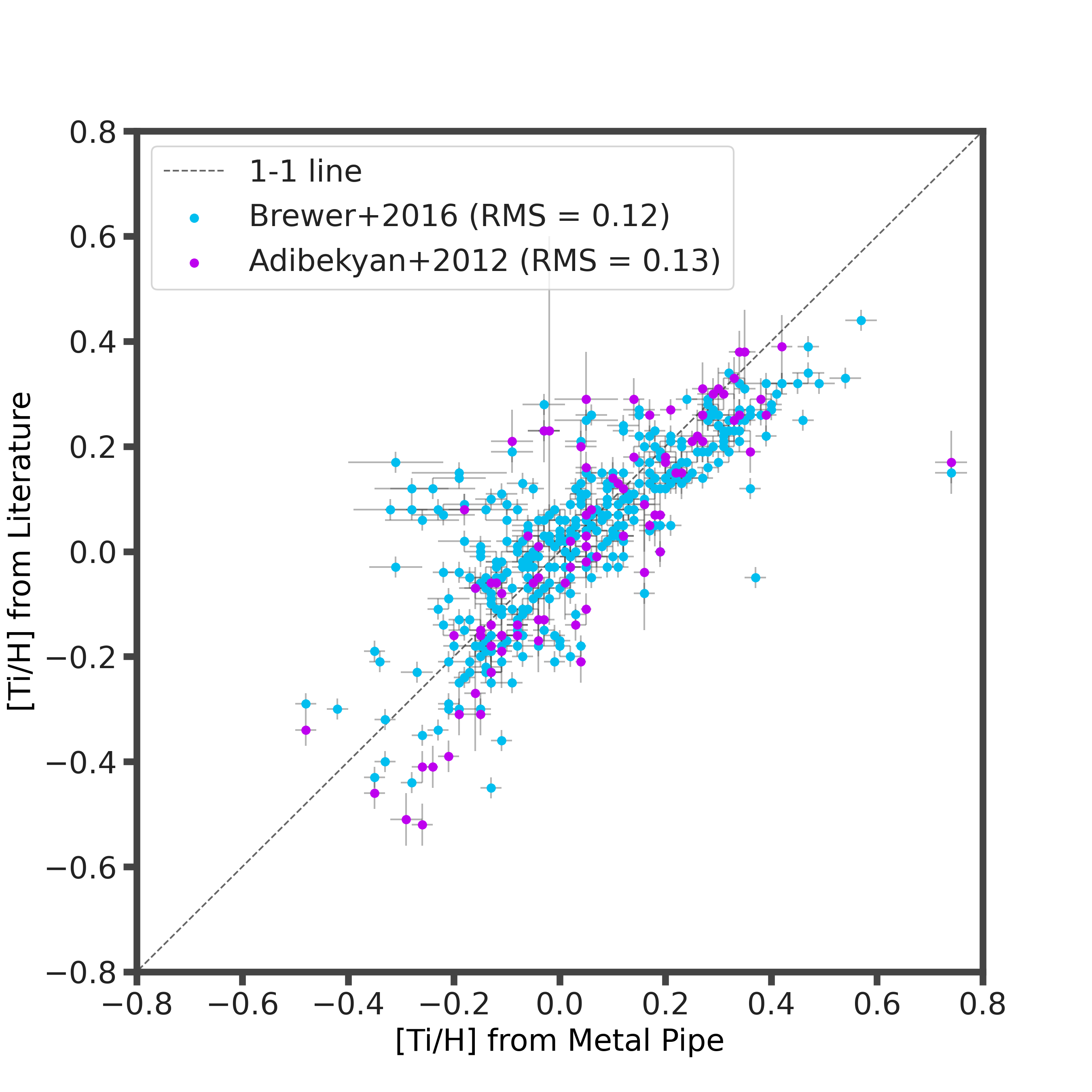}
    \caption{Comparison of titanium abundances derived by \mpp and by literature sources}
    \label{fig:22}
\end{figure*}

\begin{figure*}
    \centering
    \includegraphics[width=\linewidth]{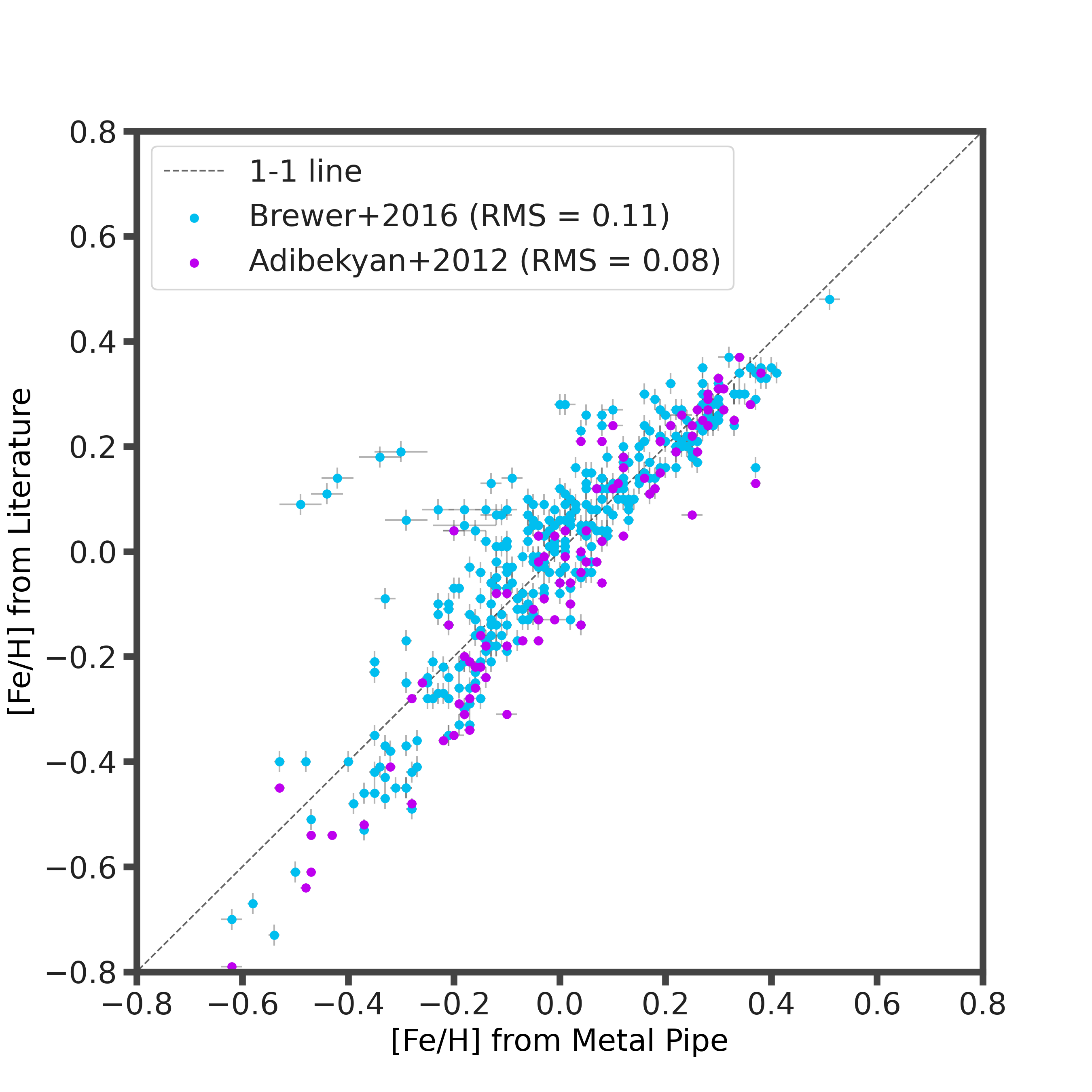}
    \caption{Comparison of iron abundances derived by \mpp and by literature sources}
    \label{fig:26}
\end{figure*}

Our results compare well with literature values \citep{Adibekyan+2012,Brewer+2016}. The RMS scatter of the literature with our work is comparable to or smaller than those found by \citet{Hinkel+2016,Petigura+2017,BC2019,Jofre+2018}. Summary statistics of our sample distributions are shown in Table \ref{table:distComp}. Figures \ref{fig:6}, \ref{fig:8}, \ref{fig:11}, \ref{fig:12}, \ref{fig:13}, \ref{fig:14}, \ref{fig:20}, \ref{fig:22}, and \ref{fig:26} show the abundance trends of C, O, Na, Mg, Al, Si, S, Ca, Ti, and Fe with literature values. Intrinsic errors reported by \mpp vary between 0.01 and 0.15 dex , and the typical RMS scatter with respect to literature values was 0.11 dex.

\begin{deluxetable*}{rccc}[t]
\tablecaption{
    Summary of methodological differences between this work and the literature works we benchmarked against.\label{table:models}
}

\tablehead{
\colhead{} &
\colhead{\textbf{This Work}} &
\colhead{\textbf{Brewer+2016}} &
\colhead{\textbf{Adibekyan+2012}}
}

\startdata
\textbf{Stellar Parameters:} & SED Fit + MIST & Spectroscopic Fit + SME  & Abundance Trends + MOOG  \\
\textbf{Model Atmospheres:} & PHOENIX  & ATLAS & ATLAS  \\
\textbf{Radiative Transfer Code:} & MOOG & SME  & MOOG \\
\textbf{Line Data Source:} & linemake + VALD & VALD* & VALD \\
\textbf{Abundance Fitting:} & Spectrum Synthesis& Spectrum Synthesis& Equivalent width matching \\
\enddata

\tablecomments{\citet{Brewer+2016} adjusted the transition strengths of lines from the values reported in the VALD database.}
\end{deluxetable*}

\begin{table}[h]
    \footnotesize
    \centering
    \begin{tabular}{|c|rr|rr|}
    \hline
      & \multicolumn{2}{c|}{\textbf{This Work - B16}} & \multicolumn{2}{c|}{\textbf{This Work - A12}} \\
      \textbf{Element} & \textbf{Median} & \textbf{Std. Dev.} & \textbf{Median} & \textbf{Std. Dev.} \\
      \hline
      C  & -0.01 & $\pm$0.09 &   --  &  --  \\
      O  &  0.02 & $\pm$0.12 &   --  &  --  \\
      Na & -0.02 & $\pm$0.08 &  0.01 & $\pm$0.10 \\
      Mg & -0.05 & $\pm$0.11 & -0.04 & $\pm$0.12 \\
      Al & -0.04 & $\pm$0.07 & -0.03 & $\pm$0.07 \\
      Si &  0.00 & $\pm$0.08 & -0.01 & $\pm$0.09 \\
      Ca &  0.03 & $\pm$0.09 & -0.01 & $\pm$0.11 \\
      Ti &  0.04 & $\pm$0.11 &  0.02 & $\pm$0.11 \\
      Fe &  0.02 & $\pm$0.11 & -0.02 & $\pm$0.08 \\
      \hline
    \end{tabular}
    \caption{Summary of our abundance distributions relative to the literature values, for stars in common with the literature source (either B16 \citep{Brewer+2016} or A12 \citep{Adibekyan+2012}).}
    \label{table:distComp}
\end{table}

\subsubsection{Abundance Outliers}
\mpp was able to retrieve abundances consistent with the literature for the vast majority of our sample. However, certain stars have evaded accurate characterization. Here we discuss these cases.

\begin{itemize}
    \item \textbf{Fast Rotators:} \mpp performs poorly on stars with rotational $v \sin{i}\gtrsim 10 \textrm{km/s}$. This may be due to the following factors: First, inaccuracies in our line broadening approximation methods may result in incorrect abundances. Additionally, rotational variability which affects the star's photometry could manifest in our analysis as poorly constrained stellar parameters. Out of over 500 stars, we noted 7 examples of fast rotators in our sample, confirmed via visual inspection of their spectra: HD 377, HD 129333, HD 132173, HD 133295, HD 152555, HD 202917, HD 210302.
    \item \textbf{Cool stars:} Elemental abundances of F and G type stars are easily measured using optical spectra. However, some K type stars are subject to some of the same effects that make analysis of M dwarfs difficult. In K dwarfs, some weaker spectral features like the O I triplet can be completely undetectable. This effect is exacerbated in metal-poor K dwarfs. Additionally, in this cooler temperature regime, molecular absorption bands are strong and numerous, creating blends with many of the lines that are otherwise good for abundance analysis. Stars in our sample which were subject to these effects include the metal-poor K-dwarfs HD 4628 (\teff = 5048 K) and HD 65277 (\teff = 4740 K).
\end{itemize}

\section{Summary and Future Prospects}\label{Sec:Future}

In this work, we have described \mpp, our new pipeline for analyzing stellar abundances. We have designed our algorithm to analyze sun-like FGK stars, evolved stars, and low-mass K- and M-dwarfs, using the same methods for each. We tested \mpp on a sample of 503 FGK stars with high-quality abundance measurements from the literature \citep{Adibekyan+2012,Brewer+2016}. We found that \mpp agrees with literature catalogs within the current precision levels of LTE analysis methods \citep{Hinkel+2016,BC2019,Jofre+2018}.

We developed and tested our code using a sample of HIRES spectra, but \mpp has the capability to analyze data from any high-resolution optical spectrograph. It is also possible to expand the curated line lists to include additional elements, or additional wavelength ranges. These extended functionalities are beyond the scope of the current work, but will be the basis of subsequent papers. 

\subsection{Improvements to \mpp}

Several improvements are likely to reduce significant measurement outliers. This will include re-working our treatment of line broadening in \mpp to be more physically motivated. Additionally, a robust exploration of covariance between stellar parameters and abundances might improve error analysis in future versions. One method of accounting for covariance, used in \citet{Kolecki+2022}, involves re-running the abundance analysis multiple times, perturbing each stellar parameter, and adding the resulting perturbations in abundance values to the error budget. However, \mpp is much more computationally complex than the methods employed by \citet{Kolecki+2022}. This makes this particular method of accounting for covariance impractical for our purposes, increasing compute time per star to a prohibitively large degree. We will explore other methods of accounting for error covariance which may be less resource-intensive.

\subsection{Future Spectral Analysis}
We will continue to apply \mpp to other FGK stars as we work to create a large, homogeneous abundance catalog of planet-hosting stars. We will be adding more visible-wavelength line lists to \mpp as we extend the range of elements which can be analyzed. Additionally, we will be curating line lists for the near-infrared wavelength range to allow \mpp to analyze M dwarfs. We will also include solar calibrations from additional spectrographs as we incorporate data from them into our catalog.
and
As our catalog of stars continues to grow, we will gain the ability to search for correlations among stellar abundance patterns and planetary system architectures.

\begin{acknowledgements}
The authors would like to thank Evan Kirby, Roman Gerasimov, and the anonymous referee for their insightful comments and feedback on this manuscript.

The authors acknowledge support from the NASA Exoplanet Research Program (grant no. 80NSSC23K0269), a NASA-Keck Data PI award (grant no. 80NSSC25K0188), and the University of Notre Dame.  

This research has made use of the SIMBAD database, operated at CDS, Strasbourg, France.
This work has made use of data from the European Space Agency (ESA) mission
{\it Gaia} (\url{https://www.cosmos.esa.int/gaia}), processed by the {\it Gaia} Data Processing and Analysis Consortium (DPAC,
\url{https://www.cosmos.esa.int/web/gaia/dpac/consortium}). Funding for the DPAC has been provided by national institutions, in particular the institutions participating in the {\it Gaia} Multilateral Agreement.
\end{acknowledgements}

\software{Astropy \citep{Astropy13,Astropy18,Astropy}, Astroquery \citep{Ginsburg+2019}, GSL \citep{Galassi_2009}, Linemake \citep{Placco+2021}, Metal Pipe \citep{MetalPipe}, MOOG \citep{Sneden1973}, Numpy \citep{Numpy}, SciPy \citep{Virtanen+2020}}

\bibliography{bib}

\end{CJK*}
\end{document}